\documentclass[superscriptaddress, twocolumn, pra,nofootinbib]{revtex4-2}
\usepackage{siunitx}
\usepackage{amsmath,amssymb}
\usepackage{graphicx}
\usepackage[pdftex,dvipsnames,usenames]{xcolor}
\usepackage[colorlinks=true,urlcolor=blue,citecolor=blue,linkcolor=blue]{hyperref}

\newcommand{\ket}[1]{\left|#1\right\rangle}
\newcommand{\bra}[1]{\left\langle #1\right|}
\DeclareMathOperator{\Tr}{Tr}

 \renewcommand{\Re}{\operatorname{{\mathrm Re}}}
  \renewcommand{\Im}{\operatorname{{\mathrm Im}}}
\newcommand{\normord}[1]{:\mathrel{#1}:}

\begin{document}
\title{Photocounting measurements with dead time and afterpulses in the continuous-wave regime}

\author{A. A. Semenov}
    \affiliation{Institut f\"ur Physik, Universit\"at Rostock, Albert-Einstein-Stra\ss{}e 23, D-18059 Rostock, Germany}
    \affiliation{Bogolyubov Institute for Theoretical Physics, NAS of Ukraine, Vul. Metrologichna 14b, 03143 Kyiv, Ukraine}
	\affiliation{Kyiv Academic University, Blvd. Vernadskogo  36, 03142  Kyiv, Ukraine}
	\affiliation{Institute of Physics, NAS of Ukraine, Prospect Nauky 46, 03028 Kyiv, Ukraine}

\author{J. Samelin}
	\affiliation{Institut f\"ur Physik, Universit\"at Rostock, Albert-Einstein-Stra\ss{}e 23, D-18059 Rostock, Germany}

\author{Ch. Boldt}
\affiliation{Institut f\"ur Physik, Universit\"at Rostock, Albert-Einstein-Stra\ss{}e 23, D-18059 Rostock, Germany}

\author{M. Sch\"unemann}
	\affiliation{Institut f\"ur Physik, Universit\"at Rostock, Albert-Einstein-Stra\ss{}e 23, D-18059 Rostock, Germany}

\author{C. Reiher}
	\affiliation{Institut f\"ur Physik, Universit\"at Rostock, Albert-Einstein-Stra\ss{}e 23, D-18059 Rostock, Germany}

\author{W. Vogel}
	\affiliation{Institut f\"ur Physik, Universit\"at Rostock, Albert-Einstein-Stra\ss{}e 23, D-18059 Rostock, Germany}	
	
\author{B. Hage}
	\affiliation{Institut f\"ur Physik, Universit\"at Rostock, Albert-Einstein-Stra\ss{}e 23, D-18059 Rostock, Germany}
    \affiliation{Department of Life, Light and Matter, Universit\"at Rostock, Albert-Einstein-Stra\ss{}e 25, D-18059 Rostock, Germany}

\date{\today}

 \begin{abstract}
	The widely used experimental technique of continuous-wave detection assumes counting pulses of photocurrent from a click-type detector inside a given measurement time window.
	With such a procedure we miss out the photons detected after each photocurrent pulse during the detector dead time.
	Additionally, each pulse may initialize so-called afterpulse, which is not associated with the real photons.
	We derive the corresponding quantum photocounting formula and experimentally verify its validity.
	Statistics of photocurrent pulses appears to be nonlinear with respect to quantum state, which is explained by the memory effect of the previous measurement time windows.
	Expressions---in general, nonlinear---connecting statistics of photons and pulses are derived for different measurement scenarios.
	We also consider an application of the obtained results to quantum state reconstruction with unbalanced homodyne detection. 
 \end{abstract}

\maketitle








\section{Introduction}
\label{Sec:Intro}
	
	Born's rule \cite{born26} lies at the heart of the measurement theory, which is an integral part of quantum physics.
	A key feature of this rule is that outcome probabilities of quantum measurements linearly depend on the density operators representing quantum states of the system, see, e.g., Refs. \cite{sinha10,harris16}.  
	Photocounting experiments conducted with quantum light are archetypal examples of such measurements.
	They are employed in many fields of fundamental and applied research such as quantum optics and quantum information.    
	
	In the most general case, the outcome of photodetectors is given by a number $n$  of clicks or photocurrent pulses usually associated with the number of detected photons.
	According to Born's rule, the probability distribution for these numbers is given by the expression
		\begin{align}\label{Eq:PhotocountingEquationHS}
			\mathcal{P}_n=\Tr\left(\hat{\rho}\,\hat{\Pi}_n\right),
		\end{align}	
	which is referred to as the photocounting formula.
	Here $\hat{\rho}$ is the density operator, characterizing the state of the given electromagnetic-field mode and $\hat{\Pi}_n$ are elements of the positive operator-valued measure (POVM), describing the detector and the measurement procedure.
	
	An alternative form of the photocounting formula (\ref{Eq:PhotocountingEquationHS}),
		\begin{align}\label{Eq:PhotocountingEquationPS}	
			\mathcal{P}_n=\int_\mathbb{C}d^2\alpha\, P(\alpha)\,\Pi_n(\alpha),
		\end{align}
	involves the Glauber-Sudarshan $P$ function $P(\alpha)$ \cite{glauber63c,sudarshan63} and  the $Q$ symbols of the POVM, $\Pi_n\left(\alpha\right)=\bra{\alpha}\hat{\Pi}_n\ket{\alpha}$, cf. Refs. \cite{cahill69,cahill69a}.
	The latter can be interpreted as the probability distribution for the numbers $n$ in the case if the electromagnetic-field mode is prepared in the coherent state $\ket{\alpha}$.
	Equation (\ref{Eq:PhotocountingEquationHS}) resembles also the photocounting formula for classical electromagnetic fields \cite{mandel_book,Mandel_1964,Lamb1969,vogel_book}.
	In this case, $P(\alpha)\geq0$ is the classical probability distribution of the complex amplitude $\alpha$ of the electromagnetic-field mode and $\Pi_n(\alpha)$ is the classical response function, being the probability to get $n$ clicks or pulses given the classical amplitude $\alpha$.
	
	In the case of photon-number-resolved (PNR) detectors, the POVM is given by the relation
		\begin{align}\label{Eq:POVM_PNR}
			\hat{\Pi}_n\equiv\hat{F}_n\left[\eta\right]=\normord{\frac{\left(\eta\hat{n}\right)^n}{n!}\exp\left(-\eta\hat{n}\right)},
		\end{align}
	cf. Refs.  \cite{mandel_book, kelley64}.
	Here $\hat{n}$ is the photon-number operator, $\eta\in[0,1]$ is the detection efficiency, and $\normord{\ldots}$ denotes normal ordering. 
	Particularly, this means that	$\hat{F}_n\left[1\right]=\ket{n}\bra{n}$ is the projector on the Fock number-state $\ket{n}$.
	The corresponding $Q$-symbols reads
		\begin{align}\label{Eq:POVM_PNR_Q}
			\Pi_n(\alpha)\equiv F_n\left[\alpha;\eta\right]
			=\frac{\left(\eta|\alpha|^2\right)^n}{n!}\exp\left(-\eta|\alpha|^2\right).
		\end{align}
	This indicates that the photon-number distribution in the case of coherent states or deterministic classical single-mode fields is the Poissonian distribution.
	
	Realistic detectors mostly cannot perfectly distinguish between adjacent numbers of  photons.
	Typical devices may only indicate their presence or absence.
	To resolve this problem at least approximately some experimental techniques have been proposed.
	An example is given by the click detectors based on spatial \cite{paul1996,castelletto2007,schettini2007,blanchet08} or temporal \cite{achilles03,fitch03,rehacek03} splitting of the light beam into several modes and then detecting each mode separately by on/off detectors.     
	The number of triggered detectors is associated with the number of photons.
	The corresponding POVM has been introduced in Ref. \cite{sperling12a}.
	Based on this consideration, the importance of the true expression for the POVM has been demonstrated. 
	For example, click statistics can be sub-Poissonian even for classical light with $P(\alpha)\geq0$.
	In this case, one can use alternative solutions for testing nonclassicality of click statistics, e.g., those proposed in Ref. \cite{sperling12c}.
	
	Another widely used experimental technique is based on counting the number of photocurrent pulses inside a measurement time window (MTW) of duration $\tau_\mathrm{m}$.
	The number of such pulses is associated with the number of photons.
	However, the direct correspondence between the number of photons and the number of pulses does not work due to a number of detector imperfections; see, e.g., Refs.~\cite{straka20,Hlousek2023}.  
	In this paper, we focus on a subset of these issues, making a significant contribution to the interpretation of the measurement results.
	First, a single pulse can correspond to several absorbed photons.
	Second, most of detectors are characterized by the dead-time interval $\tau_\mathrm{d}$.
	During this interval, which happens after each photocurrent pulse, detectors do not produce any clicks even if photons are absorbed.
	Third, detectors can produce so-called afterpulses following photon-associated pulses or other afterpulses, which are not associated with photons.
	The first issue can be easily included in the POVM in a way as it is considered for click detectors \cite{sperling12a}.
	The two last issues, however, require separate considerations.
	 
	Effects of detector dead time  on photocounting statistics of classical radiation have been widely discussed in literature; see, e.g., Refs. \cite{ricciardi66,muller73,muller74,cantor75,teich78,vannucci78, Saleh_book, rapp2019}, including the case with the presence of afterpulses \cite{straka20}.
	Standard  considerations assume employing the theory of point processes \cite{snyder_book}.
	However, a straightforward generalization of those results to the quantum case is impossible for some widely-used scenarios.
	One reason for this is that the time remaining until the end of the measurement time window (MTW) after the last pulse may be shorter than the dead-time interval. 
	This modification will affect the pulse statistics in the subsequent MTW. 
	It can be considered as a memory effect of the pulse statistics from previous MTWs. 
	Therefore, this effect must be accurately incorporated into the photodetection formula.
	Such a formula has been recently derived for superconducting nanowire single-photon detectors (SNSPDs), see Ref.~\cite{Uzunova2022}, assuming even a more general model of smooth detector recovering after each pulse.
	However, this consideration does not include the presence of afterpulses inherent to photodiodes and photomultiplier tubes. 
	
	In this paper we derive the photocounting formula, accounting for dead time and afterpulses, verify it experimentally, and demonstrate its applicability to problems of quantum optics.
	Our focus lies in providing a consistent description of the memory effect that is suitable for nonclassical states. 
	Therefore, our model is based on assumptions that restrict its applicability in the most general case, but highlight the main goal of the paper.
	First, we assume that quantum states of light are prepared for equal nonmonochromatic time modes with a rectangular envelope, which width is equal to the MTW duration $\tau_\textrm{m}$. 
	However, this may not hold true for realistic sources of nonclassical light.
	This restriction can be overcome by applying the approach presented in Refs.~\cite{Uzunova2022, Len2022} for the considered case.
	Second, we consider the detectors with nonparalyzable dead time, i.e., if a photon is absorbed during the dead-time interval, it is lost and does not affect the dead-time duration.
	Third, we assume that the detector ability to register the next photon is instantly restored after the dead time interval, which is a reasonable approximation for avalanche photodiodes but should be reconsidered for the SNSPDs \cite{Uzunova2022}.
	Fourth, we assume that the afterpulses instantly occur with a given probability after the dead-time interval.
	The last two issues are experimentally reexamined by analyzing the statistics of the time between pulses.

	We consider two main scenarios. 
	The first one, referred to as the independent MTWs, assumes darkening detector input after each MTW during the time interval exceeding $\tau_\mathrm{d}$.
	Such a procedure prevents us from the memory effect of previous MTWs.
	The same result can be reached with a proper postselection of the MTWs.  
	A theoretical description can be obtained via straightforward generalization of equations presented in   Refs. \cite{ricciardi66,muller73,muller74,cantor75,teich78,vannucci78,rapp2019,straka20}.
	However, for the sake of completeness and transparent description of further generalizations we reestablish these results also introducing a classification of measurement events on so-called regular-regular (rr) and regular-irregular (ri) parts.
	They are related to the cases when the dead-time interval of the last pulse does not exceed and exceeds, respectively, the MTW.
	         
	The second scenario, referred to as the continuous-wave (cw) detection, does not assume any darkening of detector inputs or postselections.
	Its proper consideration requires a description of irregular-regular (ir) and irregular-irregular (ii) events.
	They correspond to the cases when the dead-time interval from the previous MTW occupies a time interval at the beginning of the actual MTW.
	For our purposes it is important to get separate descriptions for the ir and ii events.
	The proper photocounting formula for the cw detection includes a nonlinear expression on the density operator $\hat{\rho}$ showing the memory effects of the previous MTWs.
	Elements of this formula can be obtained as a solution of linear recursive equations.
	We introduce its approximate solution and verify its validity with numerical simulations and experimental data.
	   
	We have theoretically considered connections between statistics of photons and  photocurrent pulses.
	In the case of cw detection this relation is nonlinear that demonstrates the memory effect of the previous MTWs.
	We have adapted the technique of geometrical reconstructions of expected values of observables \cite{kovalenko2018} to the scenario of independent MTWs.
	This is applied to quantum-state reconstruction with unbalanced homodyne detection \cite{wallentowitz96,mancini1997}. 
	   
	The paper is organized as follows:
	In Sec. \ref{Sec:Recov} we adapt the classical considerations to properly formulate the quantum photocounting formula in the scenario of independent MTWs.
	Here we also introduce the classification of the corresponding measurement events and obtain the corresponding parts of the POVM.
	A nonlinear photocounting equation in the presence of dead time and afterpulses is presented in Sec. \ref{Sec:CWD}.
	Relations between photon statistics and pulse statistics are given in Sec. \ref{Sec:Connection}.
	In Sec. \ref{Sec:Examples} we employ the obtained photocounting formula to examples of quantum states.
	Applications of the obtained results to geometrical reconstruction of expected values of observables and, particularly, to quantum-state reconstruction with unbalanced homodyne detection is considered in Sec. \ref{Sec:Geom}.
	Statistics of time passed between subsequent pulses is analyzed in Sec.~\ref{Sec:Time}. 
	The experimental verification of the photocounting formula is discussed in Sec.~\ref{Sec:Experiment}.
	A summary and concluding remarks are given in Sec. \ref{Sec:Conclusions}.


\section{Photodetection with independent measurement time windows}
\label{Sec:Recov}

\begin{figure}[ht!]
			\includegraphics[width=\linewidth]{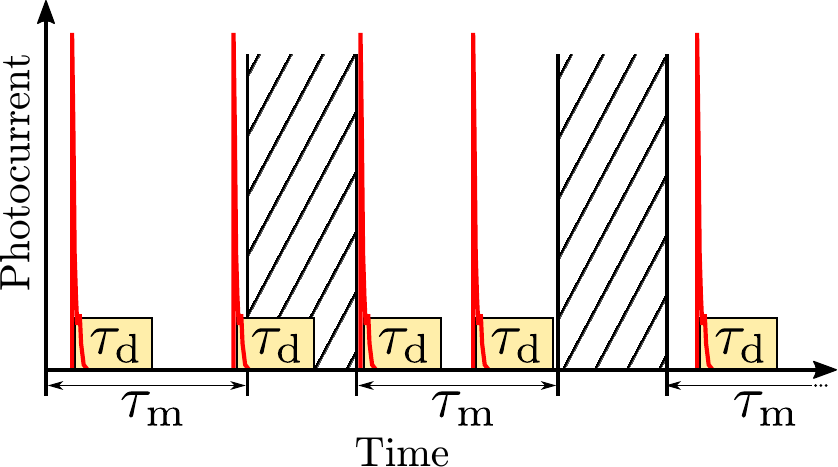}
			\caption{\label{Fig:IndTimeWind}
				The idea of photodetection with independent MTWs is depicted.
				Each MTW of duration $\tau_\mathrm{m}$ is followed by a time interval with nonoperating detector (hatched areas) exceeding the dark time $\tau_\mathrm{d}$ (shaded areas).
				In this case, the detection events in the actual MTW are not effected by the detection events from the previous one.}
		\end{figure}

	The scheme of photodetection considered in this section assumes no effect of detection events from the previous MTWs on the current one.
	It can be implemented via darkening the detector-input after each MTW.
	The corresponding time interval of darkening should exceed the dead time; cf. Fig. \ref{Fig:IndTimeWind}.  
	In practice, this purpose can be achieved in different ways, e.g., by a proper postselection of the MTWs.

	The equations obtained in this section have mostly preliminary character as an introduction to our main results.
	Most of them can be straightforwardly obtained from the corresponding classical considerations in Refs. \cite{ricciardi66,muller73,muller74,cantor75,teich78,vannucci78,rapp2019,straka20}.
	For our purposes it is, however, important to reestablish these results with a proper classification on regular-regular (rr) and regular-irregular (ri) events, which will be used by us in the next sections.

	\subsection{Dead time of detection}
	\label{Sec:DTD}
	
\begin{figure}[ht!]
			\includegraphics[width=\linewidth]{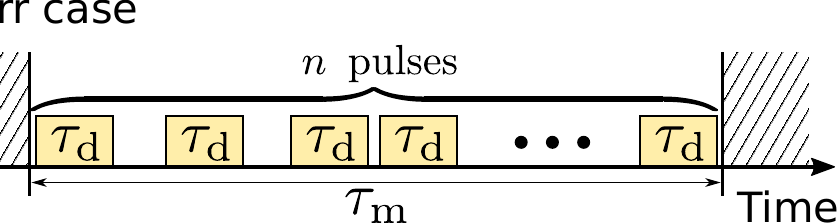}\\[4ex]
			\includegraphics[width=\linewidth]{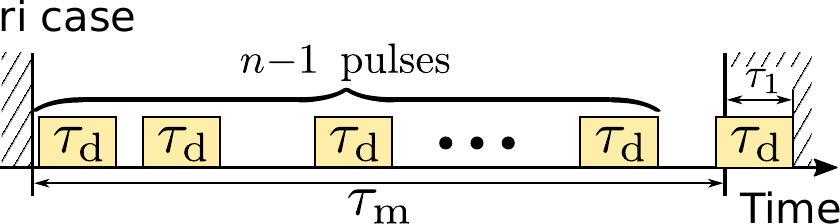}
			\caption{\label{Fig:RR-RI}
			Placements of $n$ photocurrent pulses inside the MTW for the scenario of independent MTWs is shown.		
			In the rr case, all pulses with the corresponding dead-time intervals, $\tau_\mathrm{d}$, are situated inside the MTWs.
			In the ri case, the last dead-time interval exceeds the MTW for the time $\tau_1\leq\tau_\mathrm{d}$.}
		\end{figure}

	Here we derive the photocounting formula for the case of independent MTWs assuming no afterpulses.
	For such a scenario, there exist two cases of placing the photocurrent pulses inside the time window $\tau_\mathrm{m}$, see Fig. \ref{Fig:RR-RI}.
	The regular-regular case, depicted as rr, corresponds to the case with the dead-time intervals $\tau_\mathrm{d}$ situated inside the MTW.
	The regular-irregular case, depicted as ri, corresponds to the situation wherin the last dead-time interval exceeds the time window on the time $\tau_1\leq\tau_\mathrm{d}$.
	Therefore, the POVM $\hat{\Pi}_n^\mathrm{(r)}$ can be decomposed as
		\begin{align}
			\hat{\Pi}_n^\mathrm{(r)}=\hat{\Pi}_n^\mathrm{(rr)}+\hat{\Pi}_n^\mathrm{(ri)},
			\label{Eq:POVM_IndMeas-Decomp}
		\end{align} 
	where $\hat{\Pi}_n^\mathrm{(rr)}$ and $\hat{\Pi}_n^\mathrm{(ri)}$ are the POVMs for the rr and ri cases, respectively.
	For the sake of simplicity, we first assume the unite detection efficiency and absence of dark counts.
		
	Let us start our consideration with the rr case.
	Our task is to find the $Q$ symbols of the corresponding POVM that can be interpreted as the probability of registering $n$ pulses given the coherent state $\ket{\alpha}$.
	Dead-time intervals effectively decrease the efficiency of photodetection.
	In this context, it is convenient to introduce the adjusting efficiency, which is the ratio of the time free from the dead-time intervals to the length of the MTW, 
		\begin{align}
			\eta_\mathrm{rr}(n)=\frac{\tau_\mathrm{m}-n\tau_\mathrm{d}}{\tau_\mathrm{m}}.
			\label{Eq:Eta_rr}
		\end{align}
	This quantity explicitly depends on the number of detected pulses $n$.
	The probability of registering $n$ pulses is given in the intuitively clear form by 
		\begin{align}
			\Pi_n^\mathrm{(rr)}(\alpha)=\frac{\left[\eta_\mathrm{rr}(n)|\alpha|^2\right]^n}{n!}e^{-\eta_\mathrm{rr}(n)|\alpha|^2}
			=F_n\big[\alpha;\eta_\mathrm{rr}(n)\big],
			\label{Eq:POVM_rr}
		\end{align}
	where $F_n\big(\alpha;\eta\big)$ is the $Q$ symbols of the POVM for ideal PNR detectors, cf. Eq. (\ref{Eq:POVM_PNR_Q}).
	This equation is valid for $n=0,\ldots, N$, where
		\begin{align}
			N=\left\lfloor\tau_\mathrm{m}/\tau_\mathrm{d}\right\rfloor
			\label{Eq:N}
		\end{align}
	is the maximal number of whole dead-time intervals $\tau_\mathrm{d}$ completely fitted inside the MTW and $\lfloor\ldots\rfloor$ is the floor function. 
	
	There are several ways to derive Eq.~(\ref{Eq:POVM_rr}) accurately.
	The first technique is to apply directly the methods of Ref.~\cite{Uzunova2022} for the rectangular mode.
	The second technique is based on discretizing time, applying combinatorial considerations, and transitioning back to continuous time. 
	For this purpose, we first divide the time $\tau_\mathrm{m}$ into $n_\mathrm{m}$ time slots of duration $\Delta t$, i.e., $\tau_\mathrm{m}=n_\mathrm{m}\Delta t$.
	We assume that the dead-time interval includes an integer number $n_\mathrm{d}$ of times $\Delta t$, i.e., $\tau_\mathrm{d}=n_\mathrm{d}\Delta t$.
	In this case, the probability of registering $n$ pulses consists of the following factors:
	\begin{enumerate}
		\item the probability of registering no-photons during the time free of dead-time intervals, 
		\begin{align*}
			\exp\left[-\frac{n_{\textrm{m}}-nn_{\textrm{d}}}{n_{\textrm{m}}}|\alpha|^2\right]=\exp[-\eta_\mathrm{rr}(n)|\alpha|^2];
		\end{align*}
		\item the probability of registering $n$ pulses at fixed time slots, 
		\begin{align*}
			(1-\exp[-|\alpha|^2\Delta t/\tau_\mathrm{m}])^n;
		\end{align*}
		\item the number 
		\begin{align*}
			\binom{n_\textrm{m}-nn_\textrm{d}+n}{n_\textrm{m}-nn_\textrm{d}}
		\end{align*}
		of placements $n_\textrm{m}-nn_\textrm{d}$ time slots free of dead-time intervals among $n_\textrm{m}-nn_\textrm{d}+n$ elements that contains those slots and $n$ dead-time intervals.   
	\end{enumerate}
	Applying in the product of all these factors the transition $\Delta t\rightarrow 0$, we arrive at Eq.~(\ref{Eq:POVM_rr}).
	Similar consideration can be applied in all other cases.
	Their common feature is that the probability of registering $n$ pulses can always be adjusted by the corresponding efficiency, which is the ratio of the time free of dead-time intervals to the measurement time $\tau_\mathrm{m}$.

	Let us now consider the ri case for which the last dead-time interval exceeds the MTW by the time $\tau_1$.
	The probability that the last pulse is detected during the infinitesimal time interval $d\tau_1$ is
		\begin{align}
			|\alpha|^2\frac{d\tau_1}{\tau_\mathrm{m}}\exp\left(-|\alpha|^2\frac{d\tau_1}{\tau_\mathrm{m}}\right)
			\rightarrow|\alpha|^2\frac{d\tau_1}{\tau_\mathrm{m}}.
			\label{Eq:InfProb}
		\end{align}
	This should be multiplied with the probability of registering the rest $(n-1)$ pulses in the time free from the dead-time intervals.
	The infinitesimal probability for this case is given by
		\begin{align}
			d\Pi_n^\mathrm{(ri)}&(\alpha;\tau_1)=\frac{|\alpha|^2}{\tau_\mathrm{m}}F_{n-1}[\alpha;\eta_\mathrm{ri}(\tau_1;n)]d\tau_1.
			\label{Eq:POVM_ri_Der}
		\end{align}
	Herein the adjusting efficiency reads			
		\begin{align}
			\eta_\mathrm{ri}(\tau_1;n)=\frac{\tau_\mathrm{m}-n\tau_\mathrm{d}+\tau_1}{\tau_\mathrm{m}}
		\end{align}
	and $F_k[\alpha;\eta]$ is given by Eq. (\ref{Eq:POVM_PNR_Q}).
	The resulting probability of registering $n$ pulses is obtained via the integration of Eq. (\ref{Eq:POVM_ri_Der}),
		\begin{align}
			&\Pi_n^\mathrm{(ri)}(\alpha)=\int\limits_{0}^{\tau_\mathrm{d}}d\tau_1\frac{d\Pi_n^\mathrm{(ri)}(\alpha;\tau_1)}{d\tau_1}\nonumber\\
			&=\sum\limits_{k=0}^{n-1}F_k\left[\alpha;\eta_\mathrm{rr}(n)\right]-
			\sum\limits_{k=0}^{n-1}F_k\left[\alpha;\eta_\mathrm{rr}(n-1)\right].
			\label{Eq:POVM_ri}
		\end{align}
	This equation is valid for $n=1,\ldots, N$, where $N$ is given by Eq. (\ref{Eq:N}).
	
	In the case of $\tau_\mathrm{m}/\tau_\mathrm{d}>N$, the maximal number of pulses is $N+1$.
	For this scenario, only the ri case is different from zero,
		\begin{align}
			&\Pi_{N+1}^\mathrm{(ri)}(\alpha)=\int\limits_{(N+1)\tau_\mathrm{d}-\tau_\mathrm{m}}^{\tau_\mathrm{d}}d\tau_1\frac{d\Pi_{n+1}^\mathrm{(ri)}(\alpha;\tau_1)}{d\tau_1}\nonumber\\
			&=1-
			\sum\limits_{k=0}^{N}F_k\left[\alpha;\eta_\mathrm{rr}(N)\right].
			\label{Eq:POVM_ri_Nplus1}
		\end{align}		 
	Evidently, for $\tau_\mathrm{m}=N\tau_\mathrm{d}$ this expression vanishes.
	
	The final expression for the POVM in the case of independent MTWs is obtained by substituting Eqs. (\ref{Eq:POVM_rr}), (\ref{Eq:POVM_ri}), and (\ref{Eq:POVM_ri_Nplus1}) into Eq. (\ref{Eq:POVM_IndMeas-Decomp}).
	This yields the POVM,
		\begin{align}
			\hat{\Pi}_0^\mathrm{(r)}=\hat{F}_0\left[1\right];
			\label{Eq:POVM-Ind-Meas-0}
		\end{align}
		\begin{align}
			\hat{\Pi}_n^\mathrm{(r)}
			=\sum\limits_{k=0}^{n}\hat{F}_k\left[\eta_\mathrm{rr}(n)\right]-
			\sum\limits_{k=0}^{n-1}\hat{F}_k\left[\eta_\mathrm{rr}(n-1)\right],
			\label{Eq:POVM-Ind-Meas}
		\end{align}
	for $n=1,\ldots, N$, and
		\begin{align}
		\hat{\Pi}_{N+1}^\mathrm{(r)}=1-
		\sum\limits_{k=0}^{N}\hat{F}_k\left[\eta_\mathrm{rr}(N)\right].
		\label{Eq:POVM-Ind-Meas-NP1}
		\end{align}				
	To include realistic values of detection efficiencies and the presence of dark counts, one should replace the photon-number operator $\hat{n}$ in expressions $\hat{F}_k\left[\eta_\mathrm{rr}(n)\right]$ and $\hat{F}_k\left[\eta_\mathrm{rr}(n-1)\right]$ under the sign of normal ordering, cf. Eq. (\ref{Eq:POVM_PNR}), by $\eta\hat{n}+\nu$, where $\eta$ is the detection efficiency, $\nu=\lambda_\mathrm{dc}\tau_\mathrm{m}$, and $\lambda_\mathrm{dc}$ is the dark-count rate, see, e.g., Refs.  \cite{Pratt1969,Karp1970,Lee2005,Semenov2008}. 
	Equations (\ref{Eq:POVM-Ind-Meas-0}), (\ref{Eq:POVM-Ind-Meas}), and (\ref{Eq:POVM-Ind-Meas-NP1}) can also be strightforwardly reestablished from the photocounting statistics of classical fields, see Refs.\cite{ricciardi66,muller73,muller74,cantor75,teich78,vannucci78}.
	For our further considerations, however, separate relations for $\hat{\Pi}_n^\mathrm{(rr)}$ and $\hat{\Pi}_n^\mathrm{(ri)}$ given by Eqs. (\ref{Eq:POVM_rr}), (\ref{Eq:POVM_ri}), and (\ref{Eq:POVM_ri_Nplus1}) are of importance.

	\subsection{Effect of afterpulses}
	\label{Sec:Afterpulses}
	
\begin{figure}[ht!]
				\includegraphics[width=\linewidth]{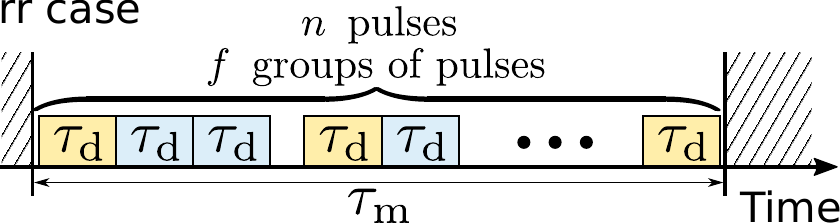}\\[4ex]
				\includegraphics[width=\linewidth]{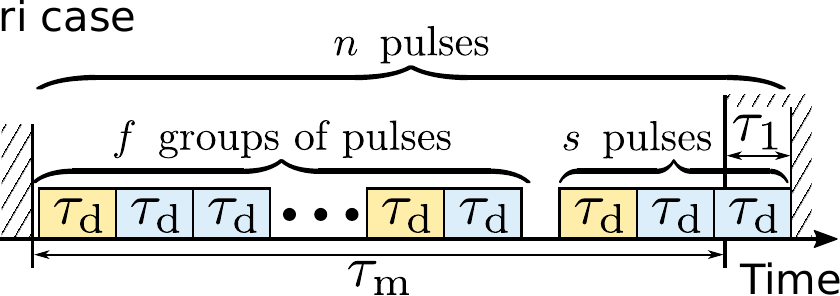}
				\caption{\label{Fig:RR-RI-AP}
					Placements of $n$ photocurrent pulses inside the MTWs in the presence of afterpulses for the scenario of independent MTWs is shown.		
					The pulses are subdivided into $f$ and $f{+}1$ groups for rr and ri cases, respectively. 
					Each group is started with a photon-related pulse and followed by afterpulses.
					The last group in the ri case consists of $s$ pulses.
					The last pulse in this group exceeds the MTW on the time $\tau_1<\tau_\mathrm{d}$.
				}
			\end{figure}

	To consider the effect of afterpulses, we use the following model.
	We assume that afterpulses may appear after each photon-related pulse, dark-count pulse, and another afterpulse with the probability $p$.
	Similar to the above scenario, we subdivide the measurement events on regular and irregular cases depicted as rr  and ri, respectively; see Fig. \ref{Fig:RR-RI-AP}.
	All pulses can be subdivided into groups started with pulses related to photons or dark counts and followed by afterpulses.
	For simplicity, we first consider the unit detection efficiency and the absence of dark counts.   
	
	Let us consider $f$ groups of pulses in the rr case.
	For the $n$ registered pulses, the number of afterpulses is $n{-}f$.
	The probability of such an event is $F_f[\alpha;\eta_\mathrm{rr}(n)]p^{n-f}(1{-}p)^f$ that includes the probability of registering $f$ photon-related pulses, the probability of registering $n{-}f$ afterpulses, and the probability of absence of $f$ afterpulses at the end of each group.
	Let $r$ be the maximal size of the groups. 
	We also consider that $f_i$ groups consist of $i$ pulses, i.e., $\sum_{i=1}^{r} f_i=f$. 
	The number of such events is given by the multinomial coefficient, i.e., by the number of all possible permutations of the groups of pulses.  
	Thus, the probability to register $n$ pulses is obtained via summation of all probabilities over all possible partitions of the number $n$:
		\begin{align}
			&\Pi_n^\mathrm{(rr)}(\alpha)\nonumber\\
			&=\sum\limits_{\sum\limits_{i=1}^{r} if_i=n}\binom{f}{f_1\ldots f_r}F_f[\alpha;\eta_\mathrm{rr}(n)]p^{n-f}(1{-}p)^f,
		\end{align}
	where $n=0\ldots N$.
	Applying the combinatorial equality
		\begin{align}\label{Eq:Combinatoric}
			\sum\limits_{\sum\limits_{i=1}^{r} if_i=n}\binom{f}{f_1\ldots f_r}A(f)=
			\sum\limits_{f=1}^{n}\binom{n-1}{f-1}A(f),
		\end{align}
	$n\neq 0$, see methods of Ref. \cite{goulden_book} for its derivation,\footnote{The authors thank E. Shchukin for enlightening discussions of this equality.} one gets the expression
		\begin{align}
			&\Pi_n^\mathrm{(rr)}(\alpha)
			&=\sum\limits_{f=1}^nF_f[\alpha;\eta_\mathrm{rr}(n)]\binom{n-1}{f-1}p^{n-f}(1{-}p)^f,
			\label{Eq:POVM_rr_ap}
		\end{align}
	which are the $Q$ symbols of the POVM for the rr case.
	The case of $n=0$ should be considered separately.
	It coincides with Eq. (\ref{Eq:POVM-Ind-Meas-0}).
	
	For the analysis of the ri case, we assume that the last group consists of $s$ pulses.
	The last pulse of this group exceeds the MTW on the time $\tau_1$.
	The rest of pulses are combined in $f$ groups, cf. Fig. \ref{Fig:RR-RI-AP}.
	The infinitesimal probability is obtained via consideration of the following factors: 
		\begin{enumerate}
			\item the probability of registering the first pulse in the last group during the infinitesimal time $d\tau_1$, cf. Eq. (\ref{Eq:InfProb});
			
			\item the probability of registering $f$ photon-related pulses, $F_f[\alpha;\eta_\mathrm{ri}(\tau_1;n)]$;
			
			\item the probability of registering $n{-}(f{+}1)=n{-}f{-}1$ afterpulses, $p^{n-f-1}$; 
			
			\item the probability of absence of afterpulses at the end of $f$ groups, $(1{-}p)^f$;
		
			\item summation for $f$ groups of $n{-}s$ pulses with the multinomial coefficient corresponding to all possible permutations;   
			
			\item summation over all possible numbers of pulses in the last group, $s=1\ldots n$.		
		\end{enumerate}
	This yields the expression,
		\begin{align}
			&d\Pi_n^\mathrm{(ri)}(\alpha;\tau_1)\nonumber\\
			&=\sum\limits_{s=1}^{n}
			\sum\limits_{\sum\limits_{i=1}^{r} if_i=n-s}\binom{f}{f_1\ldots f_r}\frac{|\alpha|^2}{\tau_\mathrm{m}}F_{n-1}[\alpha;\eta_\mathrm{ri}(\tau_1;n)]\nonumber\\
			&\times p^{n-f-1}(1{-}p)^fd\tau_1.
			\label{Eq:POVM_ri_ap_Der}
		\end{align}		   
	After proper integration with respect to $\tau_1\in[0,\tau_\mathrm{d}]$ and application of Eq. (\ref{Eq:Combinatoric}), we get
		\begin{align}
			\Pi_n^\mathrm{(ri)}(\alpha)=\sum\limits_{f=0}^{n-1}\sum\limits_{k=0}^{f}\Big\{F_k[\alpha;&\eta_\mathrm{rr}(n)]-F_k[\alpha;\eta_\mathrm{rr}(n-1)]\Big\}\nonumber\\
			&\times\binom{n-1}{f}p^{n-f-1}(1{-}p)^f.
			\label{Eq:POVM_ri_ap}
		\end{align}
	This expression is valid for $n=1\ldots N$.
	
	The special case, $n=N+1$, is obtained via integration of Eq. (\ref{Eq:POVM_ri_ap_Der}) with respect to $\tau_1\in[0,(N{+}1)\tau_\mathrm{d}{-}\tau_\mathrm{m}]$,
		\begin{align}
			&\Pi_{N+1}^\mathrm{(ri)}(\alpha)\nonumber\\
			&=1-\sum\limits_{f=0}^{N}\sum\limits_{k=0}^{f}
			F_k[\alpha;\eta_\mathrm{rr}(N)]\binom{N}{f}p^{N-f}(1{-}p)^f.
			\label{Eq:POVM_ri_ap_Nplus1}
		\end{align}
	It vanishes for $\tau_\mathrm{m}=N\tau_\mathrm{d}$.
	The rr part of the POVM in this case is zero.
	
	Combining Eqs. (\ref{Eq:POVM_IndMeas-Decomp}), (\ref{Eq:POVM_rr_ap}), (\ref{Eq:POVM_ri_ap}), and (\ref{Eq:POVM_ri_ap_Nplus1}) we arrive at the POVM for the scenario of independent MTWs in the presence of afterpulses.
	The no-click element $\hat{\Pi}_0^\mathrm{(r)}$ has the same form as Eq. (\ref{Eq:POVM-Ind-Meas-0}).
	For the rests of elements we have
		\begin{align}
			\hat{\Pi}_n^\mathrm{(r)}&=
			\sum\limits_{f=1}^n\hat{F}_f[\eta_\mathrm{rr}(n)]\binom{n-1}{f-1}p^{n-f}(1{-}p)^f\nonumber\\
			&+\sum\limits_{f=0}^{n-1}\sum\limits_{k=0}^{f}\Big\{\hat{F}_k[\eta_\mathrm{rr}(n)]-\hat{F}_k[\eta_\mathrm{rr}(n-1)]\Big\}\nonumber\\
			&\times\binom{n-1}{f}p^{n-f-1}(1{-}p)^f,
			\label{Eq:POVM-Ind-Meas-AP}
		\end{align}		
	for $n=1\ldots N$;
		\begin{align}
			\hat{\Pi}_{N+1}^\mathrm{(r)}=1-\sum\limits_{f=0}^{N}\sum\limits_{k=0}^{f}
			\hat{F}_k[\eta_\mathrm{rr}(N)]\binom{N}{f}p^{N-f}(1{-}p)^f.
			\label{Eq:POVM-Ind-Meas-NP1-AP}
		\end{align}	
	Realistic values of detection efficiencies and the presence of dark counts are included by replacing the photon-number operator $\hat{n}$ under the sign of normal ordering by $\eta\hat{n}+\nu$.
	If the probability of registering afterpulses, $p$, depends on the field intensity, the corresponding dependence $p(|\alpha|^2)$ should be explicitly included in the $Q$ symbols of the POVM.
	Equations (\ref{Eq:POVM-Ind-Meas-AP}) and (\ref{Eq:POVM-Ind-Meas-NP1-AP}) can be obtained as a straightforward generalizations of results in Ref. \cite{straka20}.
	Our further considerations, however, require a proper separation on $\hat{\Pi}_n^\mathrm{(rr)}$ and $\hat{\Pi}_n^\mathrm{(ri)}$ given by Eqs. (\ref{Eq:POVM_rr_ap}), (\ref{Eq:POVM_ri_ap}), and (\ref{Eq:POVM_ri_ap_Nplus1}).


\section{Continuous-wave detection}
\label{Sec:CWD}

	In standard implementations of cw detection, no time interruptions between the neighboring MTWs are assumed.
	If the dead-time interval from the last pulse exceeds the MTW, it changes the pulse statistics in the next MTW.
	As a result, the photocounting equation depends on the direct product of density operators from several MTWs.
	The corresponding expression is derived in this section.

	\subsection{Events dependent on previous measurements}
	
	\subsubsection{Absence of afterpulses}
	
\begin{figure}[ht!]
			\includegraphics[width=\linewidth]{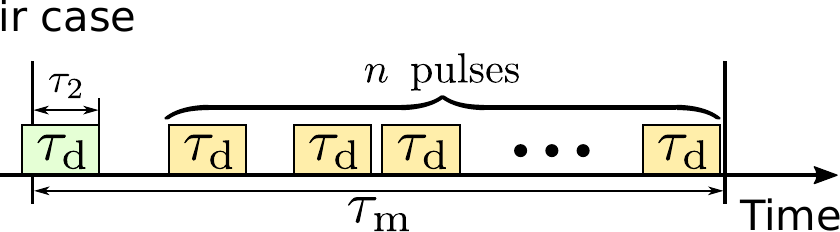}\\[4ex]
			\includegraphics[width=\linewidth]{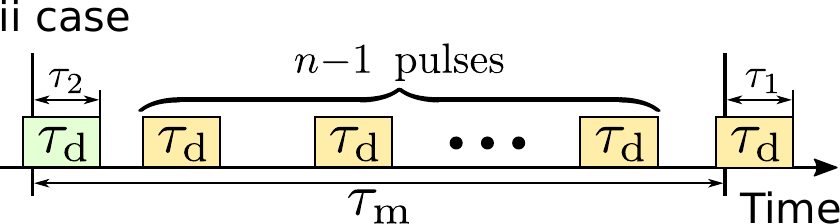}
			\caption{\label{Fig:IR-II}
				Placements of $n$ photocurrent pulses inside the MTW for the scenario of cw detection and the left-side irregular cases is shown.
				The most important difference of ir and ii cases from the cases shown in Fig. \ref{Fig:RR-RI} consists in the effect of the last pulse from the previous MTW.
				The corresponding dead-time interval occupies the time interval $\tau_2<\tau_\mathrm{d}$ at the beginning of the actual MTW.}
		\end{figure}

	First we consider the detectors with no afterpulses.
	Similar to the previous consideration, we consider irregular-regular and irregular-irregular cases depicted as ir and ii, respectively, see Fig. \ref{Fig:IR-II}.
	The both of these cases are left-side irregular in the sense that the last dead-time interval from the previous MTW occupies a time-interval $\tau_2$ at the beginning of the actual MTW.
	The parts of the POVM, which we obtain here, are conditioned by $\tau_2$.
	Integration with respect to this time will be considered in Sec. \ref{Sec:GenTheorMod} and Sec. \ref{Sec:Approx}.
	
	We remain that the $Q$ symbols of the POVM correspond to the probability of registering $n$ pulses given the coherent state $\ket{\alpha}$. 
	This probability for the ir case reads
		\begin{align}
			\Pi_n^\mathrm{(ir)}(\alpha;\tau_2)=
			F_n[\alpha;\eta_\mathrm{ir}(\tau_2;n)],
			\label{Eq:POVM_ii}
		\end{align}
	where the adjusting efficiency is given by 
		\begin{align}
			\eta_\mathrm{ir}(\tau_2;n)=\frac{\tau_\mathrm{m}-n\tau_\mathrm{d}-\tau_2}{\tau_\mathrm{m}}.
		\end{align}
	This equation is valid for $n=1,\ldots, N{-}1$ and for $n=N$ if $\tau_2<\tau_\mathrm{m}{-}N\tau_\mathrm{d}$.
	For all other cases $\Pi_n^\mathrm{(ir)}(\alpha;\tau_2)=0$.
	
\begin{figure}[ht!]
		\includegraphics[width=0.8\linewidth]{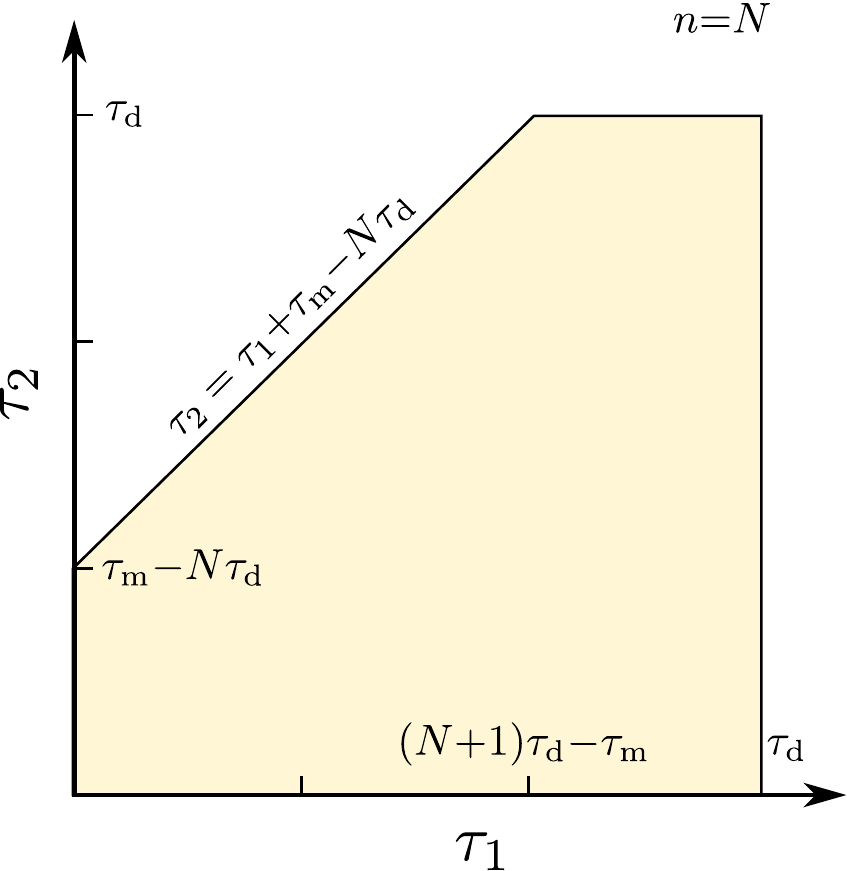}\\[4ex]
		\includegraphics[width=0.8\linewidth]{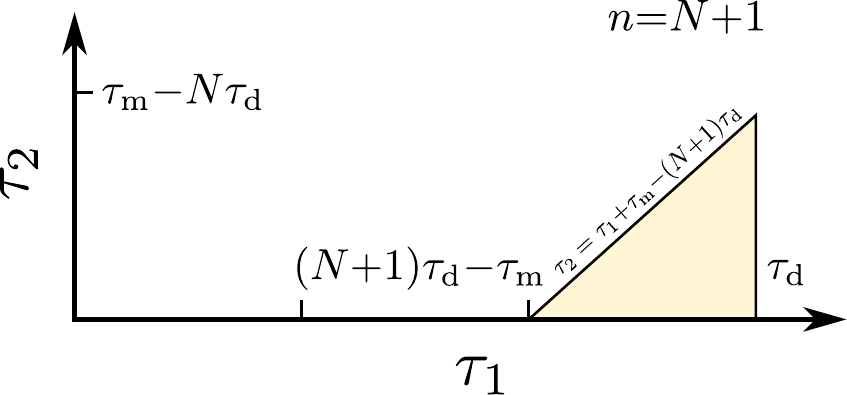}
		\caption{\label{Fig:Boarders} The shaded areas show the integration domains with respect to the times $\tau_1$ and $\tau_2$ in the  ii case for $n=N$ and $n=N+1$. 
		}
	\end{figure}

	The infinitesimal probability of registering $n$ pulses conditioned by $\tau_2$ in the ii case includes factors corresponding to the infinitesimal probability of registering the last pulse (\ref{Eq:InfProb}) and the probability of registering the rest $n{-}1$ pulses,  
		\begin{align}
			d\Pi_n^\mathrm{(ii)}(\alpha;\tau_1,\tau_2)=\frac{|\alpha|^2}{\tau_\mathrm{m}}F_{n-1}[\alpha;\eta_\mathrm{ii}(\tau_1,\tau_2;n)]d\tau_1.
			\label{Eq:POVM_ii_Der}
		\end{align}	
	Here the adjusting efficiency reads
		\begin{align}
			\eta_\mathrm{ii}(\tau_1,\tau_2;n)=\frac{\tau_\mathrm{m}-n\tau_\mathrm{d}+\tau_1-\tau_2}{\tau_\mathrm{m}}.
		\end{align}	
	This equation is valid in the domain $\tau_1,\tau_2\in[0,\tau_\mathrm{d}]$ for $n=1\ldots N{-}1$.
	For two special cases, however, there are additional conditions, see also Fig. \ref{Fig:Boarders}: 
	(i) $\tau_2-\tau_1\leq\tau_\mathrm{m}{-}N\tau_\mathrm{d}$ for $n=N$; 
	(ii) $\tau_2-\tau_1\leq\tau_\mathrm{m}{-}(N{+}1)\tau_\mathrm{d}$ for $n=N{+}1$.
	For all other cases this part of the POVM vanishes.
	Integration with respect to the time $\tau_1$ leads to the conditioned $Q$ symbol of the POVM for the ii case,
		\begin{align}
			&\Pi_n^\mathrm{(ii)}(\alpha;\tau_2)\nonumber\\
			&=\sum\limits_{k=0}^{n-1}F_k\left[\alpha;\eta_\mathrm{ir}(\tau_2;n)\right]-
			\sum\limits_{k=0}^{n-1}F_k\left[\alpha;\eta_\mathrm{ir}(\tau_2;n-1)\right]
			\label{Eq:POVM_ii_n}
		\end{align}
	for $n=1\ldots N{-}1$, $\tau_2{\in}[0,\tau_\mathrm{d}]$ and $n{=}N$, $\tau_2{\in}[0,\tau_\mathrm{m}{-}N\tau_\mathrm{d}]$;
		\begin{align}
			\Pi_N^\mathrm{(ii)}(\alpha;\tau_2)
			=1-\sum\limits_{k=0}^{N-1}F_k\left[\alpha;\eta_\mathrm{ir}(\tau_2;N-1)\right]
			\label{Eq:POVM_ii_N}
		\end{align} 
	for $\tau_2{\in}[\tau_\mathrm{m}{-}N\tau_\mathrm{d},\tau_\mathrm{d}]$;	
		\begin{align}
			\Pi_{N+1}^\mathrm{(ii)}(\alpha;\tau_2)
			=1-\sum\limits_{k=0}^{N}F_k\left[\alpha;\eta_\mathrm{ir}(\tau_2;N)\right]
			\label{Eq:POVM_ii_Nplus1}
		\end{align} 	
	for $\tau_2{\in}[0,\tau_\mathrm{m}{-}N\tau_\mathrm{d}]$.
	In all other cases this part of the POVM vanishes.
	
	The overall POVM conditioned by the time $\tau_2$ is a sum of ir and ii parts,
		\begin{align}
			\hat{\Pi}_n^\mathrm{(i)}(\tau_2)=\hat{\Pi}_n^\mathrm{(ir)}(\tau_2)+\hat{\Pi}_n^\mathrm{(ii)}(\tau_2).
			\label{Eq:POVM_Cond-Decomp}
		\end{align} 	
	Substituting here Eqs. (\ref{Eq:POVM_ii}), (\ref{Eq:POVM_ii_n}), (\ref{Eq:POVM_ii_N}), and (\ref{Eq:POVM_ii_Nplus1}) we get
		\begin{align}
			\hat{\Pi}_0^\mathrm{(i)}(\tau_2)=
			\hat{F}_0[\eta_\mathrm{(ir)}(\tau_2;0)],
		\end{align}	
		\begin{align}
			\hat{\Pi}_n^\mathrm{(i)}(\tau_2)
			=\sum\limits_{k=0}^{n}\hat{F}_k\left[\eta_\mathrm{ir}(\tau_2;n)\right]-
			\sum\limits_{k=0}^{n-1}\hat{F}_k\left[\eta_\mathrm{ir}(\tau_2;n-1)\right],
		\end{align}
	for $n=1\ldots N{-}1$, $\tau_2{\in}[0,\tau_\mathrm{d}]$ and $n{=}N$, $\tau_2{\in}[0,\tau_\mathrm{m}{-}N\tau_\mathrm{d}]$;
		\begin{align}
			\hat{\Pi}_{N}^\mathrm{(i)}(\tau_2)=1-
			\sum\limits_{k=0}^{N-1}\hat{F}_k\left[\eta_\mathrm{ir}(\tau_2;N-1)\right].
		\end{align}	
	for $\tau_2{\in}[\tau_\mathrm{m}{-}N\tau_\mathrm{d},\tau_\mathrm{d}]$;
		\begin{align}
			\hat{\Pi}_{N+1}^\mathrm{(i)}(\tau_2)=1-
			\sum\limits_{k=0}^{N}\hat{F}_k\left[\eta_\mathrm{ir}(\tau_2;N)\right].
		\end{align}			
	for $\tau_2{\in}[0,\tau_\mathrm{m}{-}N\tau_\mathrm{d}]$.
	The realistic values of the detection efficiency and the dark-count intensity are included via replacing $\hat{n}$ with $\eta\hat{n}+\nu$ under the sign of normal ordering.
	
	A special interest is attended to averaging the POVM conditioned by the time $\tau_2$ with the uniform distribution, i.e.
		\begin{align}
			\hat{\Pi}_n^{\mathrm{(i)}}=\hat{\Pi}_n^{\mathrm{(ir)}}+\hat{\Pi}_n^{\mathrm{(ii)}},
			\label{Eq:POVM_CWMeas-Decomp}
		\end{align}	
	where the corresponding $Q$ symbols are given by
		\begin{align}
			&\Pi_n^{\mathrm{(ir)}}\big(\alpha\big)=\frac{1}{\tau_\mathrm{d}}\int\limits_0^{\tau\mathrm{d}}d\tau_2\Pi_n^{\mathrm{(ir)}}(\alpha;\tau_2),
			\label{Eq:AveragingII1}\\
			&\Pi_n^{\mathrm{(ii)}}\big(\alpha\big)=\frac{1}{\tau_\mathrm{d}}\int\limits_0^{\tau\mathrm{d}}d\tau_2\Pi_n^{\mathrm{(ii)}}(\alpha;\tau_2).
			\label{Eq:AveragingII2}
		\end{align}
	The proper integration yields for $\Pi_n^{\mathrm{(ir)}}\big(\alpha\big)$
		\begin{align}
			&\Pi_n^{\mathrm{(ir)}}\big(\alpha\big)=\nonumber\\
			&\frac{\tau_\mathrm{m}}{\tau_\mathrm{d}|\alpha|^2}
			\sum\limits_{m=0}^{n}\left\{F_m\left[\alpha;\eta_\mathrm{rr}(n+1)\right]-
			F_m\left[\alpha;\eta_\mathrm{rr}(n)\right]\right\}
			\label{Eq:POVM_ir_aver}
		\end{align}
	for $n=0\ldots N-1$,
		\begin{align}
			\Pi_N^{\mathrm{(ir)}}\big(\alpha\big)=\frac{\tau_\mathrm{m}}{\tau_\mathrm{d}|\alpha|^2}
			\left\{1-\sum\limits_{m=0}^{N}F_m\left[\alpha;\eta_\mathrm{rr}(N)\right]\right\},
			\label{Eq:POVM_ir_N_aver}
		\end{align}
	for $n=N$.
	Similarly, $\Pi_n^{\mathrm{(ii)}}\big(\alpha\big)$ reads
		\begin{align}
			\Pi_n^{\mathrm{(ii)}}\big(\alpha\big)&=\frac{\tau_\mathrm{m}}{\tau_\mathrm{d}|\alpha|^2}	\sum\limits_{k=0}^{n-1}\sum\limits_{m=0}^{k}\Big\{F_m\left[\alpha;\eta_\mathrm{rr}(n+1)\right]\nonumber\\
			&-2F_m\left[\alpha;\eta_\mathrm{rr}(n)\right]+F_m\left[\alpha;\eta_\mathrm{rr}(n-1)\right]\Big\}
			\label{Eq:POVM_ii_aver}
		\end{align}
	for $n=1\ldots N-1$,
		\begin{align}
			&\Pi_N^{\mathrm{(ii)}}\big(\alpha\big)=\frac{\tau_\mathrm{m}}{|\alpha|^2\tau_\mathrm{d}}
			\left\{N-2\sum_{k=0}^{N-1}\sum_{m=0}^{k}F_m\left[\alpha;\eta_\mathrm{rr}(N)\right]\right.
			\nonumber\\
			&\left.+\sum_{k=0}^{N-1}\sum_{m=0}^{k}F_m\left[\alpha;\eta_\mathrm{rr}(N-1)\right]\right\}
			+\frac{(N+1)\tau_\mathrm{d}-\tau_\mathrm{m}}{\tau_\mathrm{d}}
			\label{Eq:POVM_ii_N_aver}
		\end{align}
	for $n=N$,
		\begin{align}
			\Pi_{N+1}^{\mathrm{(ii)}}\big(\alpha\big)=&\frac{\tau_\mathrm{m}-N\tau_\mathrm{d}}{\tau_\mathrm{d}}-\frac{\tau_\mathrm{m}(N+1)}{|\alpha|^2\tau_\mathrm{d}}\nonumber\\
			&+\frac{\tau_\mathrm{m}}{|\alpha|^2\tau_\mathrm{d}}
			\sum_{k=0}^{N}\sum_{m=0}^{k}F_m\left[\alpha;\eta_\mathrm{rr}(N)\right],
			\label{Eq:POVM_ii_Nplus1_aver}
		\end{align}
	for $n=N+1$.

	\subsubsection{Presence of afterpulses}
	
\begin{figure}[ht!]
			\includegraphics[width=\linewidth]{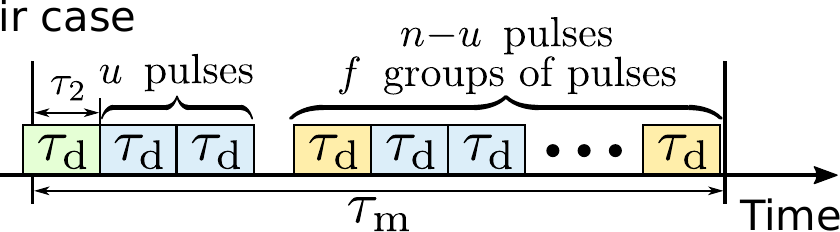}\\[4ex]
			\includegraphics[width=\linewidth]{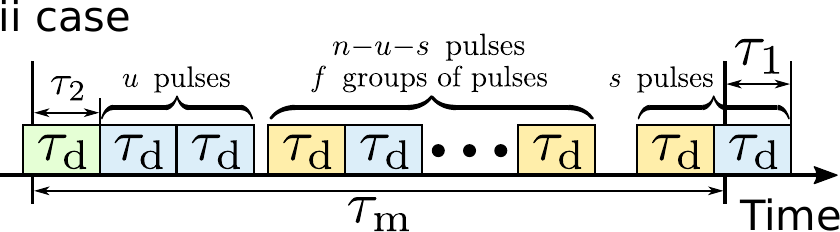}
			\caption{\label{Fig:IR-II-AP}
				Placements of $n$ photocurrent pulses inside the MTW for the scenario of cw detection and the left-side irregular cases in the presence of afterpulses is shown.
				The last (after-) pulse from the previous MTW exceeding it on the time $\tau_2$ is followed by $u$ afterpulses.
				In the ii case the last (after-) pulse in the last group of $s$ pulses exceeds the MTW on $\tau_1$.
				The rest of $n{-}u$ and $n{-}u{-}s$ in the ir and ii cases, respectively, form $f$ groups.  
			}
		\end{figure}

	In the presence of afterpulses, the situation is changed.
	In the ir and ii cases a group of $u$ afterpulses can be associated with the last (after-) pulse from the previous MTW, see Fig. \ref{Fig:IR-II-AP}.
	In the ii case the last group consists $s$ pulses.
	Therefore, the rest of $n{-}u$ and $n{-}u{-}s$ pulses for the ir and ii cases, respectively, form $f$ groups.

	The $Q$ symbols of the ir part of the POVM are obtained via considering the following factors:
	\begin{enumerate}
		\item the probability of registering $f$ photon-related pulses, $F_f[\alpha; \tau_1]$;
		
		\item the probability of registering $n{-}f$ afterpulses, $p^{n{-}f}$;
		
		\item the probability of the absence of afterpulses at the end of $f{+}1$ groups, $(1{-}p)^{f{+}1}$;
		
		\item summation for $f$ groups of $n{-}u$ pulses with the multinomial coefficient corresponding to all possible permutations;   
		
		\item summation over all possible numbers of pulses in the first group, $u=0\ldots n$.
	\end{enumerate}
	This yields the ir part of the conditional POVM,
		\begin{align}
			&\Pi_n^\mathrm{(ir)}(\alpha;\tau_2)\nonumber\\
			&=\sum\limits_{f=0}^{n}F_f[\alpha;\eta_\mathrm{ir}(\tau_2;n)]\binom{n}{f}p^{n-f}(1{-}p)^{f+1}.
			\label{Eq:POVM_ir_ap}
		\end{align}		
	This expression holds for $n=0,\ldots, N{-}1$, $\tau_2\in[0,\tau_\mathrm{d}]$  and for $n=N$, $\tau_2\in[0,\tau_\mathrm{m}{-}N\tau_\mathrm{d}]$.

	The infinitesimal conditional probability of registering $n$ pulses given the time $\tau_1$ in the ii case can be obtained via considering the following factors:
	\begin{enumerate}
	 	\item the probability of registering the first pulse in the last group during the infinitesimal time $d\tau_1$, cf. Eq. (\ref{Eq:InfProb});
	 	
	 	\item the probability of registering $f$ photon-related pulses, $F_f[\alpha;\eta_\mathrm{ii}(\tau_1,\tau_2;n)]$;
	 	
	 	\item the probability of registering $n{-}f{-}1$ afterpulses, $p^{n-f-1}$; 
	 	
	 	\item the probability of the absence of afterpulses at the end of $f{+}1$ groups, $(1{-}p)^{f+1}$;
	 	
	 	\item summation for $f$ groups of $n{-}s{-}u$ pulses with the multinomial coefficient corresponding to all possible permutations;   
 	
		\item summation over all possible numbers of pulses in the first group, $u=0\ldots n{-}s$;

	 	\item summation over all possible numbers of pulses in the last group, $s=1\ldots n$.		
	 \end{enumerate}
	This leads to the infinitesimal $Q$ symbols of the POVM in the ii case,
		\begin{align}
			&d \Pi_n^\mathrm{(ii)}(\alpha;\tau_1,\tau_2)=\frac{|\alpha|^2}{\tau_\mathrm{m}}d\tau_1\nonumber\\
			&\times\sum_{f=0}^{n-1}F_{f}[\alpha;\eta_\mathrm{ii}(\tau_1,\tau_2;n)]\binom{n}{f+1}p^{n-f-1}(1{-}p)^{f+1}.
		\end{align} 
	The $Q$ symbols of the ii part of the POVM given the time $\tau_2$ are obtained via integration of this expression with respect to $\tau_1$ in the same domains as it considered for the case without afterpulses.
	This yields
		\begin{align}
			&\Pi_n^\mathrm{(ii)}(\alpha;\tau_2)\nonumber\\
			&=\sum\limits_{f=0}^{n-1}\sum\limits_{k=0}^{f}\Big\{F_k\left[\alpha;\eta_\mathrm{ir}(\tau_2;n)\right]-
			F_k\left[\alpha;\eta_\mathrm{ir}(\tau_2;n-1)\right]\Big\}\nonumber\\
			&\times\binom{n}{f+1}p^{n-f-1}(1{-}p)^{f+1}
			\label{Eq:POVM_ii_ap}
		\end{align}
	for $n=1,\ldots, N{-}1$, $\tau_2{\in}[0,\tau_\mathrm{d}]$ and $n{=}N$, $\tau_2{\in}[0,\tau_\mathrm{m}{-}N\tau_\mathrm{d}]$;
		\begin{align}
			&\Pi_N^\mathrm{(ii)}(\alpha;\tau_2)
			=1-\sum\limits_{f=0}^{N-1}\sum\limits_{k=0}^{f}F_k\left[\alpha;\eta_\mathrm{ir}(\tau_2;N-1)\right]\nonumber\\
			&\times \binom{N}{f+1}p^{N-f-1}(1{-}p)^{f+1}
			\label{Eq:POVM_ii_N_ap}
		\end{align} 
	for $\tau_2{\in}[\tau_\mathrm{m}{-}N\tau_\mathrm{d},\tau_\mathrm{d}]$;	and 
		\begin{align}
			&\Pi_{N+1}^\mathrm{(ii)}(\alpha;\tau_2)
			=1-\sum\limits_{f=0}^{N}\sum\limits_{k=0}^{f}F_k\left[\alpha;\eta_\mathrm{ir}(\tau_2;N)\right]\nonumber\\
			&\times \binom{N+1}{f+1}p^{N-f}(1{-}p)^{f+1}
			\label{Eq:POVM_ii_Nplus1_ap}
		\end{align}
	for $\tau_2{\in}[0,\tau_\mathrm{m}{-}N\tau_\mathrm{d}]$.		
	In all other cases this part of the POVM vanishes.
	
	The overall POVM conditioned by the time $\tau_2$ is obtained from Eq. (\ref{Eq:POVM_Cond-Decomp}).
	This yields
		\begin{align}
			\hat{\Pi}_0^\mathrm{(i)}(\tau_2)=
			\hat{F}_0[\eta_\mathrm{ir}(\tau_2;0)](1-p),
		\end{align}
		\begin{align}
			\hat{\Pi}_n^\mathrm{(i)}&(\tau_2)=
			\sum\limits_{f=1}^n\hat{F}_f[\eta_\mathrm{ir}(\tau_2;n)]\binom{n}{f}p^{n-f}(1{-}p)^{f+1}\nonumber\\
			&+\sum\limits_{f=0}^{n-1}\sum\limits_{k=0}^{f}\Big\{\hat{F}_k\left[\eta_\mathrm{ir}(\tau_2;n)\right]-
			\hat{F}_k\left[\eta_\mathrm{ir}(\tau_2;n-1)\right]\Big\}\nonumber\\
			&\times\binom{n}{f+1}p^{n-f-1}(1{-}p)^{f+1}
		\end{align}		
	for $n{=}0\ldots N-1$, $\tau_2\in[0,\tau_\mathrm{m}]$ and $n{=}N$   $\tau_2{\in}[0,\tau_\mathrm{m}{-}N\tau_\mathrm{d}]$,	
		\begin{align}
			&\hat{\Pi}_N^\mathrm{(i)}(\tau_2)
			=1-\sum\limits_{f=0}^{N-1}\sum\limits_{k=0}^{f}\hat{F}_k\left[\eta_\mathrm{ir}(\tau_2;N-1)\right]\\
			&\times \binom{N}{f+1}p^{N-f-1}(1{-}p)^{f+1},
		\end{align} 	
	for $\tau_2{\in}[\tau_\mathrm{m}{-}N\tau_\mathrm{d},\tau_\mathrm{d}]$, and
		\begin{align}
			&\hat{\Pi}_{N+1}^\mathrm{(i)}(\alpha;\tau_2)
			=1-\sum\limits_{f=0}^{N}\sum\limits_{k=0}^{f}\hat{F}_k\left[\eta_\mathrm{ir}(\tau_2;N)\right]\\
			&\times \binom{N+1}{f+1}p^{N-f}(1{-}p)^{f+1}
		\end{align} 		
	for  $\tau_2{\in}[0,\tau_\mathrm{m}{-}N\tau_\mathrm{d}]$.
	Similarly to the previously considered cases, the efficiency $\eta$ and dark-count intensity $\nu$ can be included by replacing $\hat{n}$ by $\eta\hat{n}+\nu$.
	Also the dependence of probability of afterpulses on the intensity $|\alpha|^2$ should be included in the corresponding $Q$ symbols of operators.
	
	Let us consider averaging of this part of the POVM by the uniform distribution such as it is presented by Eqs. (\ref{Eq:AveragingII1}) and (\ref{Eq:AveragingII2}).
	The part $\Pi_n^{\mathrm{ir}}(\alpha)$ in the presence of afterpulses is given by
		\begin{align}
			\Pi_n^{\mathrm{ir}}(\alpha)&=\frac{\tau_\mathrm{m}}{\tau_\mathrm{d}|\alpha|^2}\sum\limits_{f=0}^{n}\binom{n}{f}p^{n-f}(1{-}p)^{f+1}\nonumber\\
			&\times\sum\limits_{m=0}^{f}\Big\{F_m[\alpha;\eta_\mathrm{rr}(n+1)]-F_m[\alpha;\eta_\mathrm{rr}(n)]\Big\}
			\label{Eq:POVM_ir_aver_ap}
		\end{align}
	for $n=0\ldots N-1$,
		\begin{align}
			\Pi_N^{\mathrm{ir}}(\alpha)=\frac{\tau_\mathrm{m}}{\tau_\mathrm{d}|\alpha|^2}&\sum\limits_{f=0}^{N}\binom{N}{f}p^{N-f}(1{-}p)^{f+1}\nonumber\\
			&\times\Big\{1-\sum\limits_{m=0}^{f}F_m[\alpha;\eta_\mathrm{rr}(N)]\Big\}
			\label{Eq:POVM_ir_N_aver_ap}
		\end{align}
	for $n=N$.
	Similarly, the part  $\Pi_n^{\mathrm{ii}}(\alpha)$ in the presence of afterpulses reads
		\begin{align}
			\Pi_n^{\mathrm{(ii)}}\big(\alpha\big)&=\frac{\tau_\mathrm{m}}{\tau_\mathrm{d}|\alpha|^2}	\sum\limits_{f=0}^{n-1}\binom{n}{f+1}p^{n-f-1}(1{-}p)^{f+1}\nonumber\\
			&\times\sum\limits_{m=0}^{f}(f-m+1)\Big\{F_m\left[\alpha;\eta_\mathrm{rr}(n+1)\right]\nonumber\\
			&-2F_m\left[\alpha;\eta_\mathrm{rr}(n)\right]
			+F_m\left[\alpha;\eta_\mathrm{rr}(n-1)\right]\Big\}
			\label{Eq:POVM_ii_aver_ap}
			\end{align}
	for $n=1\ldots N-1$,
		\begin{align}
			&\Pi_N^{\mathrm{(ii)}}\big(\alpha\big)=
			\frac{(N+1)\tau_\mathrm{d}-\tau_\mathrm{m}}{\tau_\mathrm{d}}+\frac{N\tau_\mathrm{m}}{|\alpha|^2\tau_\mathrm{d}}\left(1-p\right)\nonumber\\
			&+\frac{\tau_\mathrm{m}}{|\alpha|^2\tau_\mathrm{d}}
			\sum\limits_{f=0}^{N-1}\binom{N}{f+1}p^{N-f-1}(1{-}p)^{f+1}\nonumber\\
			&\times\sum\limits_{m=0}^{f}(f-m+1)\Big\{F_m\left[\alpha;\eta_\mathrm{rr}(N-1)\right]\nonumber\\
			&{-}2F_m\left[\alpha;\eta_\mathrm{rr}(N)\right]\Big\}
			\label{Eq:POVM_ii_N_aver_ap}
		\end{align}
		for $n=N$,	
		\begin{align}
			\Pi_{N+1}^{\mathrm{(ii)}}&\big(\alpha\big)=\frac{\tau_\mathrm{m}-N\tau_\mathrm{d}}{\tau_\mathrm{d}}-\frac{\tau_\mathrm{m}}{|\alpha|^2\tau_\mathrm{d}}(1-p)(N+1)\nonumber\\
			&+\frac{\tau_\mathrm{m}}{|\alpha|^2\tau_\mathrm{d}}\sum\limits_{f=0}^{N}\binom{N+1}{f+1}p^{N-f}(1{-}p)^{f+1}\nonumber\\
			&\times\sum_{m=0}^{f}(f-m+1)F_m\left[\alpha;\eta_\mathrm{rr}(N)\right],
			\label{Eq:POVM_ii_Nplus1_aver_ap}
		\end{align}
	for $n=N+1$.	
	
	The time-conditioned POVM $\hat{\Pi}_n^\mathrm{(i)}(\tau_2)$ and the uniformly-averaged POVM $\hat{\Pi}_n^\mathrm{(i)}$ can also be obtained as a strightforward generalization of classical photocounting statistics presented in Ref. \cite{straka20}.
	However, as it is discussed in the next section, the proper photocounting formula requires also separate relations for $\hat{\Pi}_n^\mathrm{(ii)}(\tau_2)$ and $\hat{\Pi}_n^\mathrm{(ir)}(\tau_2)$ given by Eqs. (\ref{Eq:POVM_ir_ap}), (\ref{Eq:POVM_ii_ap}), (\ref{Eq:POVM_ii_N_ap}), and (\ref{Eq:POVM_ii_Nplus1_ap}).
	Similarly, an approximation to this POVM considered in Sec. \ref{Sec:Approx} requires separate relations for $\hat{\Pi}_n^\mathrm{(ii)}$ and $\hat{\Pi}_n^\mathrm{(ir)}$ given by Eqs. (\ref{Eq:POVM_ir_aver_ap}), (\ref{Eq:POVM_ir_N_aver_ap}), (\ref{Eq:POVM_ii_aver_ap}), (\ref{Eq:POVM_ii_N_aver_ap}), and (\ref{Eq:POVM_ii_Nplus1_aver_ap}).

	\subsection{General theoretical model}
	\label{Sec:GenTheorMod}
	
	One of the main results of this paper is a special form of the photocounting formula for the cw detection.
	It differs from its standard form  (\ref{Eq:PhotocountingEquationHS}) and (\ref{Eq:PhotocountingEquationPS}) not only by another POVM but also by a different dependence on the density operator. 
	We numerate the MTWs, starting from the first one.
	Evidently, the $Q$ symbols of the multiwindow POVM, $\Lambda_n(\boldsymbol{\alpha}^{l})$, which are also interpreted as the probabilities of registering $n$ pulses, are functions of the field amplitudes in the actual and in all previous MTWs, where $\boldsymbol{\alpha}^{l}=(\alpha_1\ldots \alpha_l)$.\footnote{We indicate dimension of the string $(\alpha_1\ldots \alpha_l)$ by the upper index at the corresponding bold symbol, $\boldsymbol{\alpha}^{l}$.}
	Consequently, the photocounting formula in this case is given by
		\begin{align}
			\mathcal{P}_n=\int_{\mathbb{C}^l}d^{2l}\boldsymbol{\alpha}\, P(\alpha_l) P(\alpha_{l-1})\ldots P(\alpha_1)\,\Lambda_n\big(\boldsymbol{\alpha}^{l}\big).
			\label{Eq:PhCountEq_Full_P}
		\end{align}
	The same equation rewritten in the operator form reads
		\begin{align}\label{Eq:PhotocountingEquationGeneral}
			\mathcal{P}_n=\Tr\left(\hat{\rho}^{\otimes l}\,\hat{\Lambda}_n\right).
		\end{align}
	This photoconting equation includes a nonlinear dependence on the density operator $\hat{\rho}$ that is caused by the memory effect from the previous MTWs.
	
	The nonlinear dependence on the density operator can be seen as a significant modification of Born's rule. At first glance, it appears to contradict the fundamental Gleason theorem \cite{Gleason1957} in its generalized form \cite{Busch2003, Caves2004}. However, this is not actually the case, as Eq.~(\ref{Eq:PhotocountingEquationGeneral}) can still be considered a linear Born's rule with respect to the density operator $\hat{\rho}^{\otimes l}$ in the Hilbert space, which represents the tensor product of $l$ Hilbert spaces.
	 
	Let $\mathcal{Q}_l\big(\boldsymbol{\alpha}^{l-1}\big)$ and $\varrho_l\big(\tau_2;\boldsymbol{\alpha}^{l-1}\big)$ be the probability of the event with $\tau_2=0$ and the probability distribution of $\tau_2$ at the $l$th MTW, respectively.
	These quantities are normalized as
		\begin{equation}
			\mathcal{Q}_l\big(\boldsymbol{\alpha}^{l-1}\big)+\int\limits_{0}^{\tau_\mathrm{d}}d \tau_2\, \varrho_l\big(\tau_2;\boldsymbol{\alpha}^{l-1}\big)=1.
			\label{Eq:Normalization}
		\end{equation} 
	Evidently, they also depend on the field amplitudes $\boldsymbol{\alpha}^{l-1}$ of previous $l{-}1$ MTWs. 
	The $Q$ symbols of the multiwindow POVM are given by  
		\begin{align}
			\Lambda_n\big(\boldsymbol{\alpha}^{l}\big)=&\mathcal{Q}_l\big(\boldsymbol{\alpha}^{l-1}\big)\Pi_n^{\mathrm{(r)}}(\alpha_l)\nonumber\\
			&+\int\limits_0^{\tau\mathrm{d}}d\tau_2\varrho_l\big(\tau_2;\boldsymbol{\alpha}^{l-1}\big)\Pi_n^{\mathrm{(i)}}(\alpha_l;\tau_2).
			\label{Eq:POVM_Gen}
		\end{align}
	This relation is obtained via averaging the r and i parts of the POVM at the $l$th MTW with the time $\tau_2$.
	
	The probability $\mathcal{Q}_l\big(\boldsymbol{\alpha}^{l-1}\big)$ and the probability distribution $\varrho_l\big(\tau_2;\boldsymbol{\alpha}^{l-1}\big)$ obey the recurrence relations, which can be derived from rules of the probability theory,
		\begin{align}
			\mathcal{Q}_{l+1}\big(\boldsymbol{\alpha}^l\big)=\mathcal{Q}_{l}&\big(\boldsymbol{\alpha}^{l-1}\big)A(\alpha_l)\nonumber\\
			&+\int\limits_0^{\tau_\mathrm{d}}d \tau \varrho_l\big(\tau;\boldsymbol{\alpha}^{l-1}\big)B(\alpha_l;\tau),
			\label{Eq:Recc1}
		\end{align} 
		\begin{align}
			\varrho_{l+1}\big(\tau;\boldsymbol{\alpha}^{l}\big)=&\mathcal{Q}_{l}\big(\boldsymbol{\alpha}^{l-1}\big)G(\alpha_l;\tau)\nonumber\\
			&+\int\limits_0^{\tau_\mathrm{d}}d \tau_2 \varrho_l\big(\tau^\prime;\boldsymbol{\alpha}^{l-1}\big)H(\alpha_l;\tau^\prime,\tau).
			\label{Eq:Recc2}
		\end{align} 
	Here 
		\begin{align}\label{Eq:Al}
			A(\alpha_l)=\sum\limits_{n=0}^{N}{\Pi}_{n}^\mathrm{(rr)}(\alpha_l)
		\end{align}
	is the probability of occurring the rr case in the $l$th MTW,
		\begin{align}\label{Eq:Bl}
			B(\alpha_l;\tau)=\sum\limits_{n=0}^{N}{\Pi}_{n}^\mathrm{(ir)}(\alpha_l;\tau)
		\end{align}
	is the probability of occurring the ir case in the $l$th MTW given the last-pulse exceedance time $\tau$ in the $(l-1)$th MTW,
		\begin{align}
			G(\alpha_l;\tau)=\sum\limits_{n=1}^{N+1}{\Pi}_{n}^\mathrm{(ri)}(\alpha_l;\tau)
		\end{align} 
	is the probability density of the last-pulse exceedance time $\tau$ for the ri case in the $l$th MTW,
		\begin{align}
			H(\alpha_l;\tau^\prime,\tau)=\sum\limits_{n=1}^{N+1}{\Pi}_{n}^\mathrm{(ii)}(\alpha_l;\tau,\tau^\prime)
		\end{align} 
	is the probability density of the last-pulse exceedance time $\tau$ for the ii case in the $l$th MTW given the last-pulse exceedance time $\tau^\prime$ in the $(l-1)$th MTW.
	Hence, $\mathcal{Q}_l\big(\boldsymbol{\alpha}^{l-1}\big)$ and $\varrho\big(\tau_2;\boldsymbol{\alpha}^{l-1}\big)$ for any MTW can be obtained via applying the recurrence procedure given by Eqs. (\ref{Eq:Recc1}) and (\ref{Eq:Recc2}) with the initial conditions
		\begin{align}
			\mathcal{Q}_1=1,\quad \varrho_1(\tau_2)=0.
		\end{align} 
	
	Technically, the obtained quantities and, as a consequence, the POVM $\hat{\Lambda}_n$ depend on all previous MTWs.
	The POVM has also a specific form for each number $l$ of the MTW.
	However, as we demonstrate below, this dependence quickly disappears such that only few previous MTWs play the role.
	As a result, the POVMs for different numbers $l$ coincide with good accuracy.
	Therefore, the pulse statistics of the MTW with the number $l\gg 1$ can be obtained in a standard way via sampling of events in all MTW.

	\subsection{Approximation by the uniform distribution}	
	\label{Sec:Approx}

	Resolving the recurrence equations (\ref{Eq:Recc1}) and (\ref{Eq:Recc2}) requires involved resources.
	The task can be drastically simplified for the case of $\tau_\mathrm{d}\ll \tau_\mathrm{m}$ and light fields with intensities $|\alpha|^2\ll\tau_\mathrm{m}/\tau_\mathrm{d}$.
	In this case $\varrho(\tau_2;\boldsymbol{\alpha}^{l-1})$ can be approximately considered as the uniform distribution, i.e.,
		\begin{align}\label{Eq:RhoUniform}
			\varrho_l(\tau_2;\boldsymbol{\alpha}^{l-1})=\frac{1-\mathcal{Q}_l(\boldsymbol{\alpha}^{l-1})}{\tau_\mathrm{d}},	
		\end{align}
	which satisfies the normalization requirements (\ref{Eq:Normalization}).

	This approximation enables to obtain the multiwindow POVM in a simple form.  
	Substituting Eq. (\ref{Eq:RhoUniform}) in Eq. (\ref{Eq:POVM_Gen}) we arrive at the corresponding expression,
		\begin{align}
			\Lambda_n\big(\boldsymbol{\alpha}^{l}\big)=\mathcal{Q}_l&\big(\boldsymbol{\alpha}^{l-1}\big)\Pi_n^{\mathrm{(r)}}(\alpha_l)\nonumber\\
			&+\left[1-\mathcal{Q}_l\big(\boldsymbol{\alpha}^{l-1}\big)\right]\Pi_n^{\mathrm{(i)}}\big(\alpha_l\big),
			\label{Eq:POVM_Appr}
		\end{align}
	where $\Pi_n^{\mathrm{(r)}}(\alpha_l)$ and $\Pi_n^{\mathrm{(i)}}(\alpha_l)$ are given by Eqs.  (\ref{Eq:POVM_IndMeas-Decomp}) and (\ref{Eq:POVM_CWMeas-Decomp}), respectively.
	The corresponding components are given by Eq. (\ref{Eq:POVM_rr_ap}) for $\Pi_n^{\mathrm{(rr)}}(\alpha_l)$, Eqs. (\ref{Eq:POVM_ri_ap}) and (\ref{Eq:POVM_ri_ap_Nplus1}) for $\Pi_n^{\mathrm{(ri)}}(\alpha_l)$,  Eqs. (\ref{Eq:POVM_ir_aver_ap}) and (\ref{Eq:POVM_ir_N_aver_ap}) for $\Pi_n^{\mathrm{(ir)}}(\alpha_l)$, and  Eqs. (\ref{Eq:POVM_ii_aver_ap}), (\ref{Eq:POVM_ii_N_aver_ap}), and (\ref{Eq:POVM_ii_Nplus1_aver_ap}) for $\Pi_n^{\mathrm{(ii)}}(\alpha_l)$ in the presence of afterpulses.
	In the absence of afterpulses these quantities are reduced to Eq. (\ref{Eq:POVM_rr}) for $\Pi_n^{\mathrm{(rr)}}(\alpha_l)$, Eqs. (\ref{Eq:POVM_ri}) and (\ref{Eq:POVM_ri_Nplus1}) for $\Pi_n^{\mathrm{(ri)}}(\alpha_l)$,  Eqs. (\ref{Eq:POVM_ir_aver}) and (\ref{Eq:POVM_ir_N_aver}) for $\Pi_n^{\mathrm{(ir)}}(\alpha_l)$, and  Eqs. (\ref{Eq:POVM_ii_aver}), (\ref{Eq:POVM_ii_N_aver}), and (\ref{Eq:POVM_ii_Nplus1_aver}) for $\Pi_n^{\mathrm{(ii)}}(\alpha_l)$.
	
	As the next step, we obtain the recurrence relation for the probability $\mathcal{Q}_l(\boldsymbol{\alpha}^{l-1})$.
	For this purpose, we substitute Eq. (\ref{Eq:RhoUniform}) in Eq. (\ref{Eq:Recc1}).
	This yields 
	 	\begin{align}
			 \mathcal{Q}_{l+1}\big(\boldsymbol{\alpha}^l\big)=\mathcal{Q}_{l}&\big(\boldsymbol{\alpha}^{l-1}\big)C(\alpha_l)+B(\alpha_l),
			 \label{Eq:ReccQ}
		 \end{align} 
	where
		\begin{align}
			B(\alpha_l)&=\frac{1}{\tau_\mathrm{d}}\int\limits_0^{\tau\mathrm{d}}d\tau B(\alpha_l;\tau)
			=\sum\limits_{n=0}^{N}{\Pi}_{n}^\mathrm{(ir)}(\alpha_l),
		\end{align}
		\begin{align}\label{Eq:Cl}
			C(\alpha_l)=A(\alpha_l)-B(\alpha_l),
		\end{align}
	and $A(\alpha_l)$ given by Eq. (\ref{Eq:Al}).
	The recurrence relation (\ref{Eq:ReccQ}) is resolved as 
		\begin{align}
			\mathcal{Q}_{l}\big(\boldsymbol{\alpha}^{l-1}\big)=&B(\alpha_{l-1})+B(\alpha_{l-2})C(\alpha_{l-1})\nonumber\\
			&+B(\alpha_{l-3})C(\alpha_{l-2})C(\alpha_{l-1})+\ldots. 
			\label{Eq:ProbabilityQ}
		\end{align}
	The structure of this equation has a clear interpretation.
	The first term describes influence of events from the last MTW.
	The second and subsequent terms describe influence of events from the penultimate and earlier MTWs, respectively.
	In fact, the contribution of such terms vanishes with increasing the number of the involved MTWs.
	For many practical cases, it is enough to consider only two first terms.
	This means that the considered process is ergodic to a good approximation. 
	This implies that statistics of each MTW can be obtained by averaging statistics from all MTWs except the first ones.


\section{Pulse statistics vs photon statistics}
\label{Sec:Connection}

	A special interest is attended to the Fock-state representation of the POVM $\hat{\Pi}_n^{\mathrm{(r)}}$.
	For an arbitrary photocounting procedure this results in the expansion
		\begin{align}
			\hat{\Pi}_n^{\mathrm{(r)}}=\sum\limits_{k=0}^{+\infty}P^{\mathrm{(r)}}_{n|k}\ket{k}\bra{k},
			\label{Eq:FockRepr}
		\end{align}
	where $P^{\mathrm{(r)}}_{n|k}=\bra{k}\hat{\Pi}_n^{\mathrm{(r)}}\ket{k}$ is the probability of registering $n$ clicks (pulses) given $k$ photons, $\ket{k}$ is the Fock number state; see, e.g., discussions in Refs. \cite{sperling12a,kovalenko2018,Len2022}.
	In the absence of extra events such as dark counts or afterpulses, this probability satisfies the clear condition $P_{n|k}=0$ for $n>k$.   
	This means that summation in Eq. (\ref{Eq:FockRepr}) starts in this case from $n$.

	To separate the effects solely related to the imperfect resolution of photon numbers caused by dead time of detection and afterpulses, we consider the expansion by the POVM $\hat{F}_n[\eta]$ given by Eqs. (\ref{Eq:POVM_PNR}) and (\ref{Eq:POVM_PNR_Q}).
	This yields the expansion,
		\begin{align}
			\hat{\Pi}_n^{\mathrm{(r)}}=\sum\limits_{k=0}^{+\infty}P^{\mathrm{(r)}}_{n|k}\hat{F}_k[\eta],
			\label{Eq:POVMRepr}
		\end{align}		
	where $P^{\mathrm{(r)}}_{n|k}$ is now interpreted as the probability to register $n$ pulses under the condition that the detector with the efficiency $\eta$ has absorbed $k$ photons.
	Similarly, one can include in this consideration dark counts.
	By averaging Eq. (\ref{Eq:POVMRepr}) with the density operator $\hat{\rho}$ we obtain the relation,
		\begin{align}
			\mathcal{P}_n=\sum\limits_{k=0}^{+\infty}P^{\mathrm{(r)}}_{n|k}P_k[\eta],
			\label{Eq:PulseStatVSPhotStat1}
		\end{align}
	between the probability to get $n$ pulses, $\mathcal{P}_n$, and the probability to absorb $k$ photons by the detector with the efficiency $\eta$, $P_k[\eta]$.
	  
	In the case of cw detection, wherin the events from previous time windows affect on the events from the current one, Eq. (\ref{Eq:POVMRepr}) is replaced by its nonlinear form,
		\begin{align}
			\hat{\Pi}_n=\sum\limits_{k_1,\ldots,k_l=0}^{+\infty}\Lambda_{n|k_{l},\ldots,k_1}\hat{F}_{k_l}[\eta]\otimes
			\ldots\otimes\hat{F}_{k_1}[\eta].
			\label{Eq:POVMReprCW}
		\end{align}		
	Here $\Lambda_{n|k_{l},\ldots,k_1}=\bra{k_l,\ldots,k_1}\hat{\Lambda}_n\ket{k_l,\ldots,k_1}$ is the probability of registering $n$ pulses given $k_{l},k_{l-1}\ldots,k_2,k_1$ absorbed photons in the MTWs with the numbers $l,l{-}1,\ldots,2,1$, respectively. 
	By averaging this relation with $\hat{\rho}^{\otimes l}$, we arrive at the expression
		\begin{align}
			\mathcal{P}_n=\sum\limits_{k_1,\ldots,k_l=0}^{+\infty}\Lambda_{n|k_{l},\ldots,k_1}P_{k_l}[\eta]
			\ldots P_{k_1}[\eta].
			\label{Eq:PulseStatVSPhotStat2}
		\end{align}	
	This expression states the nonlinear relation between the probability of registering $n$ pulses, $\mathcal{P}_n$, and the probability to absorb $k$ photons by the detector with the efficiency $\eta$, $P_k[\eta]$.
	
	\subsection{Independent MTWs}
	
	First, we consider the case of independent MTWs described by Eq. (\ref{Eq:PulseStatVSPhotStat1}).
	To find the probabilities $P^{\mathrm{(r)}}_{n|k}$ one can expand the $Q$ symbols of the POVM, $\Pi_n^{\mathrm{(r)}}(\alpha)$, by the $Q$ symbols of the projectors $\ket{k}\bra{k}$ given by $F_n[\alpha;\eta=1]$.
	Applying Eq. (\ref{Eq:POVM_IndMeas-Decomp}) we obtain the decomposition,
		\begin{align}
			P_{n|k}^{\mathrm{(r)}}=P_{n|k}^{\mathrm{(rr)}}+P_{n|k}^{\mathrm{(ri)}},
			\label{Eq:Pr_Decomposition}
		\end{align}
	where $P_{n|k}^{\mathrm{(rr)}}=\bra{k}\hat{\Pi}_{n}^{\mathrm{(rr)}}\ket{k}$ and $P_{n|k}^{\mathrm{(ri)}}=\bra{k}\hat{\Pi}_{n}^{\mathrm{(ri)}}\ket{k}$. 
	
	In the absence of afterpulses we get for the rr part
		\begin{align}
			P_{n|k}^{\mathrm{(rr)}}=\binom{k}{n}\left[\eta_\mathrm{rr}(n)\right]^n
			\left[1-\eta_\mathrm{rr}(n)\right]^{k-n},
			\label{Eq:Prr}
		\end{align}
	for $n=0\ldots N$.
	The ri part reads
		\begin{align}
			P_{n|k}^{\mathrm{(ri)}}=&\sum\limits_{m=k-n+1}^{k}\binom{k}{m}
			\Big\{\left[\eta_\mathrm{rr}(n)\right]^{k-m}
			\left[1-\eta_\mathrm{rr}(n)\right]^{m}\nonumber\\
			&-\left[\eta_\mathrm{rr}(n-1)\right]^{k-m}
			\left[1-\eta_\mathrm{rr}(n-1)\right]^{m}\Big\}
			\label{Eq:Pri}
		\end{align}	
	for $n=0\ldots N$,
		\begin{align}
			&P_{N+1|k}^{\mathrm{(ri)}}\nonumber\\&=1-\sum\limits_{m=k-N}^{k}\binom{k}{m}
			\left[\eta_\mathrm{rr}(N)\right]^{k-m}
			\left[1-\eta_\mathrm{rr}(N)\right]^{m}
			\label{Eq:Pri_Nplus1}
		\end{align}			
	for $n=N+1$.
	As mentioned, $P_{n|k}^{\mathrm{(rr)}}=0$ for $n>k$. 
    A particular form of these equations in the case of $n=k$ has also been considered in Ref.~\cite{Len2022}.
	
	In the presence of afterpulses the rr part is given by 
		\begin{align}
			P_{n|k}^{\mathrm{(rr)}}=\sum\limits_{f=1}^{\min(k,n)}\binom{k}{f}&
			\left[\eta_\mathrm{rr}(n)\right]^f
			\left[1-\eta_\mathrm{rr}(n)\right]^{k-f}\nonumber\\
			&\times\binom{n-1}{f-1}p^{n-f}(1-p)^f.
			\label{Eq:Prr_ap}
		\end{align}
	The ri part in this case reads
		\begin{align}
			P_{n|k}^{\mathrm{(ri)}}=\sum\limits_{f=0}^{\min(k-1,n-1)}\binom{n-1}{f}p^{n-f-1}(1-p)^f\nonumber\\
			\times\sum\limits_{m=k-f}^{k}\binom{k}{m}
			\Big\{\left[\eta_\mathrm{rr}(n)\right]^{k-m}
			\left[1-\eta_\mathrm{rr}(n)\right]^{m}\nonumber\\
			-\left[\eta_\mathrm{rr}(n-1)\right]^{k-m}
			\left[1-\eta_\mathrm{rr}(n-1)\right]^{m}\Big\}
			\label{Eq:Pri_ap}
		\end{align}	
	for $n=0\ldots N$,
		\begin{align}
			P_{N+1|k}^{\mathrm{(ri)}}&=1-\sum\limits_{f=k}^{N}\binom{N}{f}p^{N-f}(1-p)^f\nonumber\\
			&-\sum\limits_{f=0}^{\min(N,k-1)}\binom{N}{f}p^{N-f}(1-p)^f\nonumber\\
			&\times\sum\limits_{m=k-f}^{k}\binom{k}{m}
			\left[\eta_\mathrm{rr}(N)\right]^{k-m}
			\left[1-\eta_\mathrm{rr}(N)\right]^{m}
			\label{Eq:Pri_Nplus1_ap}
		\end{align}			
	for $n=N+1$.	
	The second term of the last equation takes zero value for $k>N$.
	In this scenario, the conditional probabilities $P_{n|k}\neq0$ even for $n>k$ that is caused by contribution of afterpulses.
	
	\subsection{Nonlinear relation for the cw detection}
	
	The conditional probability $\Lambda_{n|k_{l},\ldots,k_1}$ included in Eqs. (\ref{Eq:POVMReprCW}) and (\ref{Eq:PulseStatVSPhotStat2}) can be obtained as the Fock-state representation of Eq. (\ref{Eq:POVM_Appr}).
	We suppose that the uniform-distribution approximation is applicable.
	This yields the decomposition,
		\begin{align}
			\Lambda_{n|k_{l},\ldots,k_1}=
			\left[P_{n|k_l}^{\mathrm{(r)}}-P_{n|k_l}^{\mathrm{(i)}}\right]Q_{k_{l-1}\ldots k_1}
			+P_{n|k_l}^{\mathrm{(i)}}.
			\label{Eq:CondProb_CW}
		\end{align}
	Here $P_{n|k}^{\mathrm{(r)}}$ is given by Eq. (\ref{Eq:Pr_Decomposition}), similarly
		\begin{align}
			P_{n|k}^{\mathrm{(i)}}=P_{n|k}^{\mathrm{(ir)}}+P_{n|k}^{\mathrm{(ii)}},
			\label{Eq:Pi_Decomposition}
		\end{align}		
	where $P_{n|k}^{\mathrm{(ir)}}=\bra{k}\hat{\Pi}_{n}^{\mathrm{(ir)}}\ket{k}$ and $P_{n|k}^{\mathrm{(ii)}}=\bra{k}\hat{\Pi}_{n}^{\mathrm{(ii)}}\ket{k}$ are Fock-state representation of the averaged parts of the POVM given by Eqs. (\ref{Eq:AveragingII1}) and (\ref{Eq:AveragingII2}), respectively.
	Employing Eq. (\ref{Eq:ProbabilityQ}) we obtain
		\begin{align}
			Q_{k_{l-1}\ldots k_1}=B_{k_{l-1}}&+B_{k_{l-2}}C_{k_{l-1}}\nonumber\\
			&+B_{k_{l-3}}C_{k_{l-2}}C_{k_{l-1}}+\ldots,
		\end{align}
	where 
		\begin{align}
			C_k=A_k-B_k,
		\end{align}
		\begin{align}
			A_k=\sum\limits_{n=0}^{N}P_{n|k}^\mathrm{(rr)},\quad B_k=\sum\limits_{n=0}^{N}P_{n|k}^\mathrm{(ir)}.
		\end{align}
	Therefore, all constituents of Eq. (\ref{Eq:CondProb_CW}) are expressed in terms of $P_{n|k}^\mathrm{(rr)}$, $P_{n|k}^\mathrm{(ri)}$ [see Eqs. (\ref{Eq:Prr}), (\ref{Eq:Pri}), (\ref{Eq:Pri_Nplus1}) and Eqs. (\ref{Eq:Prr_ap}), (\ref{Eq:Pri_ap}), (\ref{Eq:Pri_Nplus1_ap}) for the cases of absence and presence of afterpulses, respectively] as well as $P_{n|k}^\mathrm{(ir)}$, $P_{n|k}^\mathrm{(ii)}$, which are obtained via expansion of the averaged parts of the POVM, $\hat{\Pi}_n^\mathrm{(ir)}$ and $\hat{\Pi}_n^\mathrm{(ii)}$ [see Eqs. (\ref{Eq:AveragingII1}) and (\ref{Eq:AveragingII2}), respectively], by the Fock-state projectors $\ket{k}\bra{k}$.  
	
	In the absence of afterpulses, the conditional probabilities $P_{n|k}^\mathrm{(ir)}$ are given by
		\begin{align}
			P_{n|k}^\mathrm{(ir)}&=\frac{\tau_\mathrm{m}}{\tau_\mathrm{d}}\binom{k}{n}
			B_{n\frac{\tau_\mathrm{d}}{\tau_\mathrm{m}},(n+1)\frac{\tau_\mathrm{d}}{\tau_\mathrm{m}}}\left(k-n+1,n+1\right)
		\end{align}
	for $n{=}0\ldots N-1$,
		\begin{align}
			P_{N|k}^\mathrm{(ir)}&=\frac{\tau_\mathrm{m}}{\tau_\mathrm{d}}\binom{k}{N}
			B_{N\frac{\tau_\mathrm{d}}{\tau_\mathrm{m}},1}\left(k-N+1,N+1\right)
		\end{align}	
	for $n{=}N$.	
	The conditional probability $P_{n|k}^\mathrm{(ii)}$ reads 
		\begin{align}
			P_{n|k}^\mathrm{(ii)}=\frac{\tau_\mathrm{m}}{\tau_\mathrm{d}}&\nonumber\\
			\times
			\sum\limits_{l=k-n+1}^{k}
			&\binom{k}{l}\left[B_{n\frac{\tau_\mathrm{d}}{\tau_\mathrm{m}},(n+1)\frac{\tau_\mathrm{d}}{\tau_\mathrm{m}}}\left(l+1,k-l+1\right)\right.\nonumber\\
			&\left.-B_{(n-1)\frac{\tau_\mathrm{d}}{\tau_\mathrm{m}},n\frac{\tau_\mathrm{d}}{\tau_\mathrm{m}}}\left(l+1,k-l+1\right)\right]
		\end{align}
	for $n{=}0\ldots N-1$,
		\begin{align}
			&P_{N|k}^\mathrm{(ii)}=N+1-\frac{\tau_\mathrm{m}}{\tau_\mathrm{d}}\nonumber\\
			&+\frac{\tau_\mathrm{m}}{\tau_\mathrm{d}}
			\sum\limits_{l=k-N+1}^{k}
			\binom{k}{l}\left[B_{N\frac{\tau_\mathrm{d}}{\tau_\mathrm{m}},1}\left(l+1,k-l+1\right)\right.\nonumber\\
			&\left.-B_{(N-1)\frac{\tau_\mathrm{d}}{\tau_\mathrm{m}},N\frac{\tau_\mathrm{d}}{\tau_\mathrm{m}}}\left(l+1,k-l+1\right)\right]
		\end{align}	
	for $n{=}N$,
		\begin{align}
			&P_{N+1|k}^\mathrm{(ii)}=\frac{\tau_\mathrm{m}}{\tau_\mathrm{d}}-N\nonumber\\
			&-\frac{\tau_\mathrm{m}}{\tau_\mathrm{d}}
			\sum\limits_{l=k-N}^{k}
			\binom{k}{l}B_{N\frac{\tau_\mathrm{d}}{\tau_\mathrm{m}},1}\left(l+1,k-l+1\right).
		\end{align}		
	Here $B_{z_1,z_2}(a,b)=\int_{z_1}^{z_2}d t\, t^{a-1}(1-t)^{b-1}$ is the incomplete generalized beta function.
	Moreover, as it has been discussed above, $P_{n|k}^\mathrm{(ii)}=0$ for $n>k$. 
	
	In the presence of afterpulses, the conditional probabilities $P_{n|k}^\mathrm{(ir)}$ read
		\begin{align}
			P_{n|k}^\mathrm{(ir)}=\frac{\tau_\mathrm{m}}{\tau_\mathrm{d}}&
			\sum\limits_{f=0}^{\min(k,n)}\binom{k}{f}\binom{n}{f}p^{n-f}(1-p)^{f+1}\nonumber\\
			&\times
			B_{n\frac{\tau_\mathrm{d}}{\tau_\mathrm{m}},(n+1)\frac{\tau_\mathrm{d}}{\tau_\mathrm{m}}}\left(k-f+1,f+1\right)
		\end{align}	
	for $n{=}0\ldots N-1$ and
		\begin{align}
			P_{N|k}^\mathrm{(ir)}=\frac{\tau_\mathrm{m}}{\tau_\mathrm{d}}
			\sum\limits_{f=0}^{\min(k,N)}&\binom{k}{f}\binom{N}{f}p^{N-f}(1-p)^{f+1}\nonumber\\
			&\times
			B_{N\frac{\tau_\mathrm{d}}{\tau_\mathrm{m}},1}\left(k-f+1,f+1\right)
		\end{align}		
	for $n{=}N$.
	The conditional probabilities $P_{n|k}^\mathrm{(ii)}$ are given by
		\begin{align}
			&P_{n|k}^\mathrm{(ii)}=\frac{\tau_\mathrm{m}}{\tau_\mathrm{d}}\nonumber\\
			&\times\sum\limits_{f=0}^{\min(k-1,n-1)}\sum\limits_{l=0}^{f}\binom{k}{l}\binom{n}{f+1}p^{n-f-1}(1-p)^{f+1}\nonumber\\
			&\times 
			\left[B_{n\frac{\tau_\mathrm{d}}{\tau_\mathrm{m}},(n+1)\frac{\tau_\mathrm{d}}{\tau_\mathrm{m}}}\left(k-l+1,l+1\right)\right.\nonumber\\
			&\left.-B_{(n-1)\frac{\tau_\mathrm{d}}{\tau_\mathrm{m}},n\frac{\tau_\mathrm{d}}{\tau_\mathrm{m}}}\left(k-l+1,l+1\right)\right]
		\end{align}	
	for $n{=}0\ldots N-1$,
		\begin{align}
			P_{N|0}^\mathrm{(ii)}=\frac{(N+1)\tau_\mathrm{d}-\tau_\mathrm{m}}{\tau_\mathrm{d}}p^N
		\end{align}
	for $n{=}N$ and $k{=}0$,
		\begin{align}
			&P_{N|k}^\mathrm{(ii)}=\frac{(N+1)\tau_\mathrm{d}-\tau_\mathrm{m}}{\tau_\mathrm{d}}\nonumber\\
			&\times\left[1-\sum\limits_{f=k}^{N-1}\binom{N}{f+1}p^{N-f-1}(1-p)^{f+1}\right]
			\nonumber\\
			&+\frac{\tau_\mathrm{m}}{\tau_\mathrm{d}}
			\sum\limits_{f=0}^{k-1}\sum\limits_{l=0}^{f}
			\binom{k}{l}\binom{N}{f+1}p^{N-f-1}(1-p)^{f+1}\nonumber\\
			&\times\left[B_{N\frac{\tau_\mathrm{d}}{\tau_\mathrm{m}},1}\left(k-l+1,l+1\right)\right.\nonumber\\
			&\left.-B_{(N-1)\frac{\tau_\mathrm{d}}{\tau_\mathrm{m}},N\frac{\tau_\mathrm{d}}{\tau_\mathrm{m}}}\left(k-l+1,l+1\right)\right]
		\end{align}
	for $n{=}N$ and $k{=}1\ldots N-1$,
		\begin{align}
			&P_{N|k}^\mathrm{(ii)}=\frac{(N+1)\tau_\mathrm{d}-\tau_\mathrm{m}}{\tau_\mathrm{d}}\nonumber\\
			&+\frac{\tau_\mathrm{m}}{\tau_\mathrm{d}}
			\sum\limits_{f=0}^{N-1}\sum\limits_{l=0}^{f}
			\binom{k}{l}\binom{N}{f+1}p^{N-f-1}(1-p)^{f+1}\nonumber\\
			&\times\left[B_{N\frac{\tau_\mathrm{d}}{\tau_\mathrm{m}},1}\left(k-l+1,l+1\right)\right.\nonumber\\
			&\left.-B_{(N-1)\frac{\tau_\mathrm{d}}{\tau_\mathrm{m}},N\frac{\tau_\mathrm{d}}{\tau_\mathrm{m}}}\left(k-l+1,l+1\right)\right]
		\end{align}	
	for $n{=}N$ and $k\geq N$;
		\begin{align}
			P_{N+1|0}^\mathrm{(ii)}=\frac{\tau_\mathrm{m}-N\tau_\mathrm{d}}{\tau_\mathrm{d}}p^{N+1}
		\end{align}
	for $n{=}N+1$ and $k{=}0$,	
		\begin{align}
			&P_{N+1|k}^\mathrm{(ii)}=\frac{\tau_\mathrm{m}-N\tau_\mathrm{d}}{\tau_\mathrm{d}}\nonumber\\
			&\times\left[1-\sum\limits_{f=k}^{N}\binom{N+1}{f+1}p^{N-f}(1-p)^{f+1}\right]
			\nonumber\\
			&-\frac{\tau_\mathrm{m}}{\tau_\mathrm{d}}
			\sum\limits_{f=0}^{k-1}\sum\limits_{l=0}^{f}
			\binom{k}{l}\binom{N+1}{f+1}p^{N-f}(1-p)^{f+1}\nonumber\\
			&\times B_{N\frac{\tau_\mathrm{d}}{\tau_\mathrm{m}},1}\left(k-l+1,l+1\right)
		\end{align}	
	for $n{=}N+1$ and $k{=}1\ldots N$, and
		\begin{align}
			&P_{N+1|k}^\mathrm{(ii)}=\frac{\tau_\mathrm{m}-N\tau_\mathrm{d}}{\tau_\mathrm{d}}\nonumber\\
			&-\frac{\tau_\mathrm{m}}{\tau_\mathrm{d}}
			\sum\limits_{f=0}^{N}\sum\limits_{l=0}^{f}
			\binom{k}{l}\binom{N+1}{f+1}p^{N-f}(1-p)^{f+1}\nonumber\\
			&\times B_{N\frac{\tau_\mathrm{d}}{\tau_\mathrm{m}},1}\left(k-l+1,l+1\right)
		\end{align}		
	for $n{=}N+1$ and $k\geq N+1$.


\section{Examples}
\label{Sec:Examples}

\begin{figure}[ht!]
		\includegraphics[width=\linewidth]{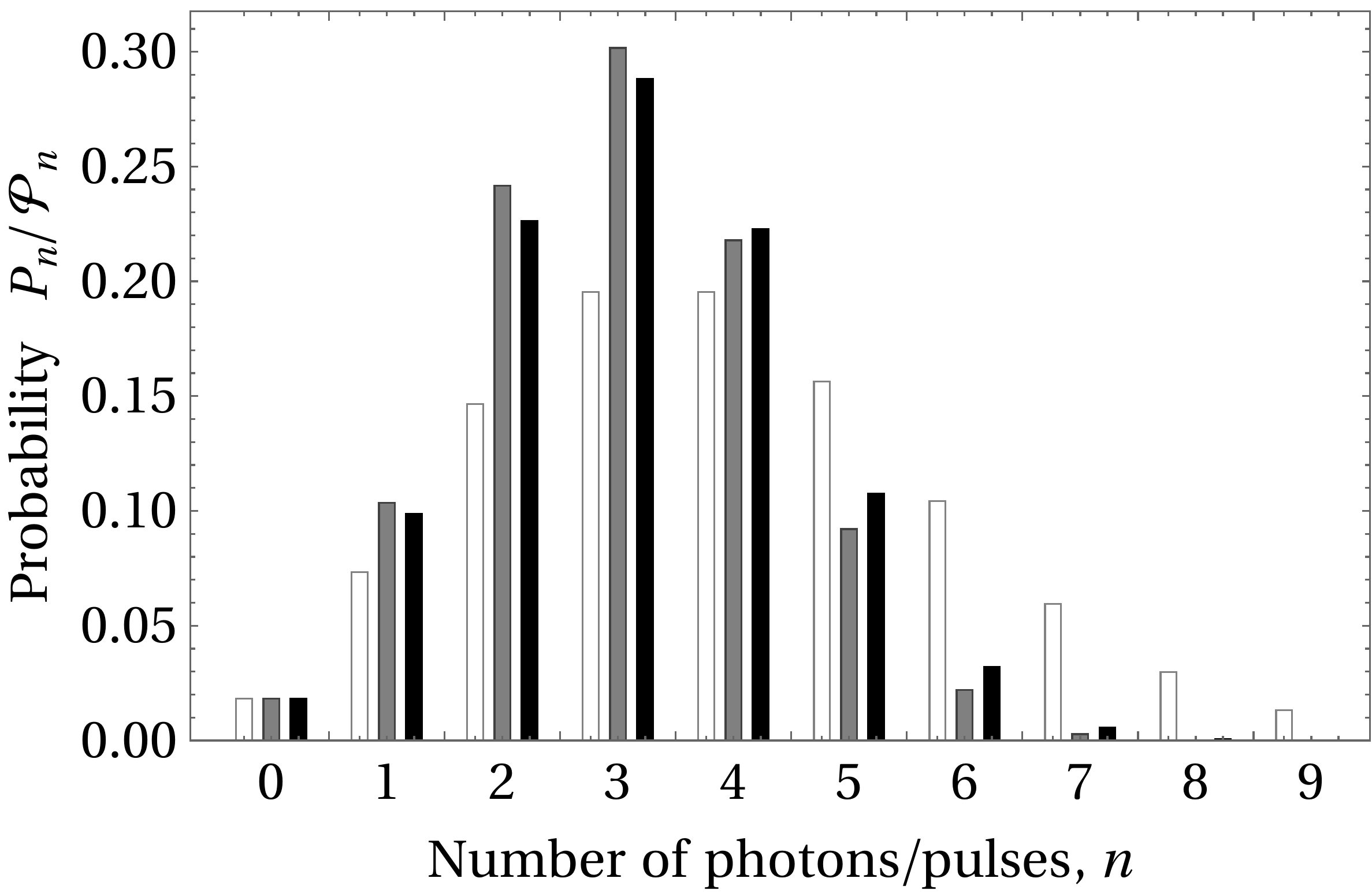}
		\caption{\label{Fig:CS} The photon-number and pulse-number distributions for coherent states $\ket{\alpha_0}$, $\alpha_0=2$, are shown.
		White bars correspond to the photon-number distribution, $P_n$.
		The gray and black bars correspond to the pulse-number distributions, $\mathcal{P}_n$, for the detection with independent MTWs without and with afterpulses, respectively. 
		Here $\tau_\mathrm{d}=0.09\tau_\mathrm{m}$, $p=0.05$.}
	\end{figure}

	In this section, we apply the derived photocounting equations to examples of quantum states of the radiation field in order to demonstrate that the finite dead time, afterpulses, and the memory effect from the previous MTWs may significantly change photocounting statistics.
	Let us start with consideration of coherent states $\ket{\alpha_0}$ given by the Glauber-Sudarshan $P$ function
		\begin{align}
			P\left(\alpha\right)=\delta\left(\alpha-\alpha_0\right).\label{Eq:PFunc_CS}
		\end{align}
	The photonumber distribution in this case is given by the Poissonian distribution,
		\begin{align}
			P_n=F_n\left[\alpha_0,1\right],
		\end{align} 
	where $F_n\left[\alpha;\eta\right]$ is given by Eq.	(\ref{Eq:POVM_PNR_Q}).
	Pulse-number distribution in the case of independent MTWs reads
		\begin{align}
			\mathcal{P}_n=\Pi_n^\mathrm{(r)}\left(\alpha_0\right),
		\end{align}
	cf. Eq.	\ref{Eq:POVM_IndMeas-Decomp}, where $Q$ symbols of the POVM are obtained from Eqs. (\ref{Eq:POVM-Ind-Meas-0}), (\ref{Eq:POVM-Ind-Meas}), and (\ref{Eq:POVM-Ind-Meas-NP1}) in the absence of afterpulses and by Eqs.(\ref{Eq:POVM-Ind-Meas-0}), (\ref{Eq:POVM-Ind-Meas-AP}), and (\ref{Eq:POVM-Ind-Meas-NP1-AP}) in the presence of afterpulses.
	These statics are compared in Fig. \ref{Fig:CS}.
	It is clearly seen that the dead time and the presence of afterpulses significantly modify this statistics.
	
	The present analysis also demonstrates that the Mandel $Q_\textrm{M}$-parameter \cite{mandel79} cannot be directly applied for testing nonclassicality with the considered measurements, cf. also Ref.~\cite{sperling12c}.
	Indeed, a naive approach assumes that the number of pulses corresponds to the number of detected photons.
	However, this leads to $Q_\textrm{M}=-0.44$ and $Q_\textrm{M}=-0.41$ in the absence and presence of afterpulses, respectively, for the cases considered in Fig. \ref{Fig:CS}.
	Therefore, direct applications of the Mandel $Q$-parameter falsely indicates nonclassicality for classical light.    

\begin{figure}[ht!]
			\includegraphics[width=\linewidth]{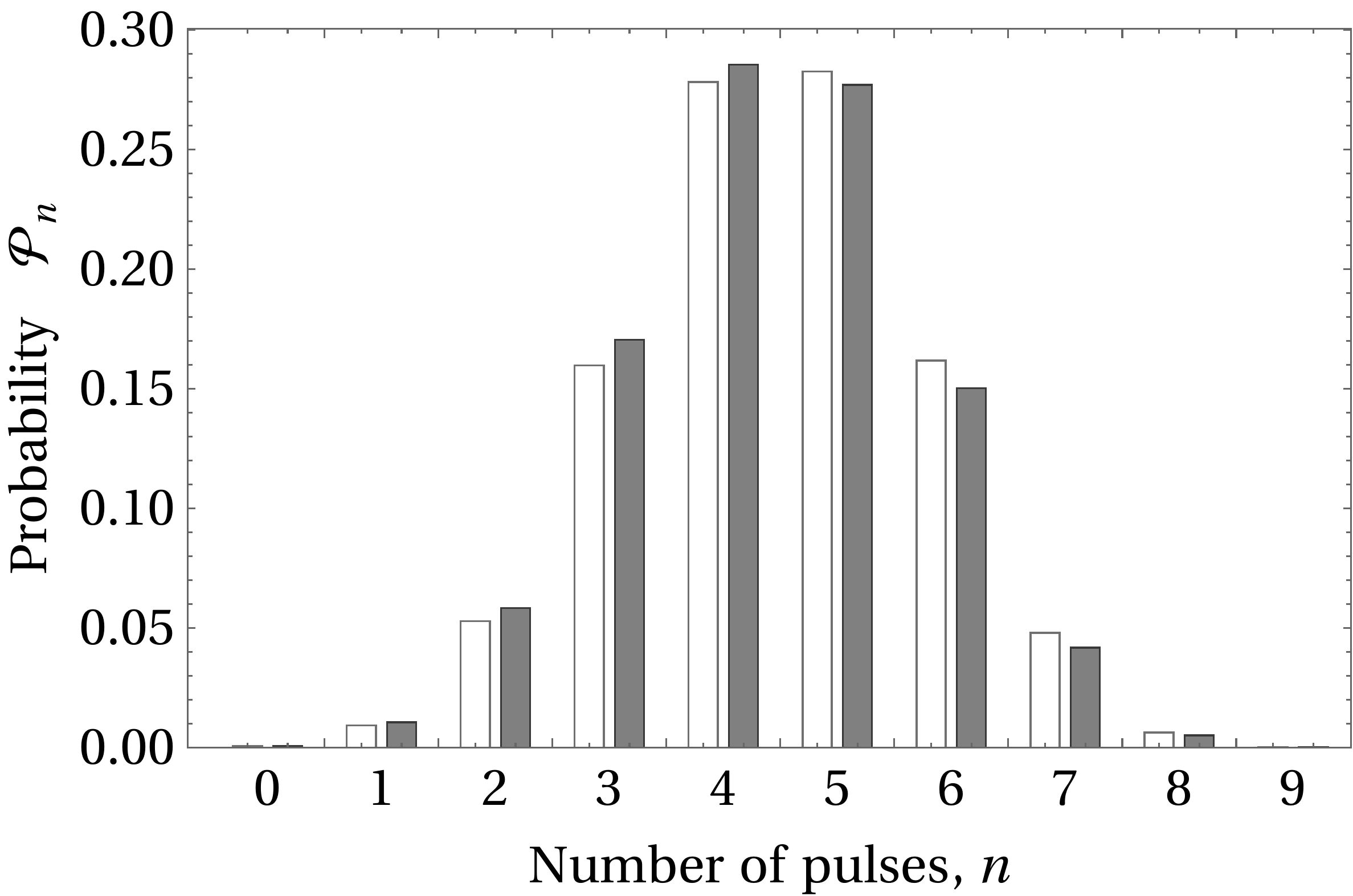}
			\caption{\label{Fig:CSMem} The pulse-number distributions, $\mathcal{P}_n$, for the  detection with independent MTWs and for the cw detection of the coherent state $\ket{\alpha_0}$, $\alpha_0=2.7$, are presented by white and gray bars, respectively.
			Here $\tau_\mathrm{d}=0.09\tau_\mathrm{m}$, $p=0$.}
		\end{figure}

	Consider the modifications of pulse statistics by the memory effect from the previous MTW.
	Substituting Eq. (\ref{Eq:PFunc_CS}) into Eq. (\ref{Eq:PhCountEq_Full_P}) and applying the approximate expression for the POVM symbols (\ref{Eq:POVM_Appr}), we arrive at the probability distribution 
		\begin{align}\label{Eq:CW_CS}
			\mathcal{P}_n=\Lambda_n(\boldsymbol{\alpha}_0^l),
		\end{align}
	where all components of $\boldsymbol{\alpha}_0^l$ are equal to $\alpha_0$.
	In Fig.~\ref{Fig:CSMem} we compare the corresponding statistics between two measurement scenarios: the detection with independent MTWs and the cw detection.
	The difference between these cases can also be clearly seen via comparison of expected values and variances: $\langle n\rangle\approx4.48$, $\langle \Delta n^2\rangle\approx1.71$ and  $\langle n\rangle\approx4.40$, $\langle \Delta n^2\rangle\approx1.69$ for the independent MTWs and the cw detection, respectively. 
	
\begin{figure}[ht!]
			\includegraphics[width=\linewidth]{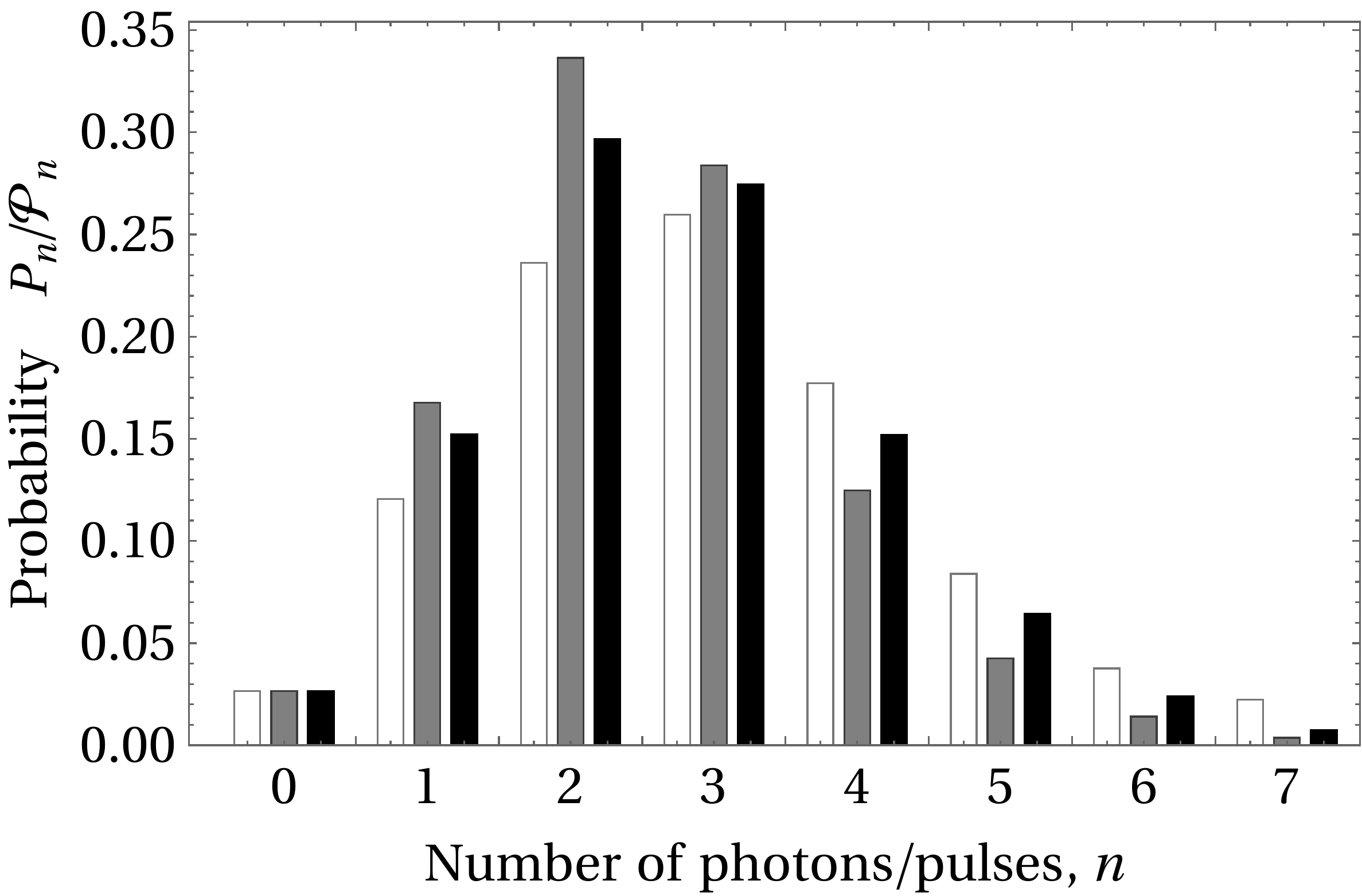}
			\caption{\label{Fig:SqCS} The photon-number and pulse-number distributions for the attenuated squeezed coherent states given by Eq. (\ref{Eq:PhSqCS}), $r=1$, $\eta=0.6$, $\alpha^\prime_0=2$, are shown.
			White bars correspond to the photon-number distribution, $P_n$.
			The gray and black bars correspond to the pulse-number distributions, $\mathcal{P}_n$, for the detection with independent MTWs without and with afterpulses, respectively. 
			Here $\tau_\mathrm{d}=0.09\tau_\mathrm{m}$, $p=0.1$.}
		\end{figure}

	Another example is a phase-squeezed coherent state given by the Wigner function
		\begin{align}
			W\left(\alpha\right)=\frac{2}{\pi}\exp\left[-2\left(\alpha^\prime-\alpha^\prime_0\right)^2e^{2r}-2\alpha^{\prime\prime 2}e^{-2r}\right],\label{Eq:PhSqCS}
		\end{align} 
	where $\alpha^\prime=\Re\alpha$, $\alpha^{\prime\prime}=\Im\alpha$, $\alpha^\prime_0$ is the coherent displacement amplitude, and $r$ is the squeezing parameter.
	The photon-number distribution $P_n$ in this case is obtained as the coefficients in the power series
		\begin{align}
			P(t)=\sum\limits_{n=0}^{+\infty}P_n t^n,
		\end{align} 
	where	
		\begin{align}
			&P(t)=\frac{1}{\sqrt{1+[\eta(2-\eta)-2t\eta(1-\eta)-t^2\eta^2]\sinh^2r}}\nonumber\\
			&\times\exp\left[\frac{\alpha^{\prime 2}\eta(1-t)[\eta(1-t)(1-e^{-2r})-2]}
			{2\Big[1+[\eta(2-\eta)-2t\eta(1-\eta)-t^2\eta^2]\sinh^2r\Big]}\right]
		\end{align}	
	and $\eta$ is the detection efficiency.
	To obtain the corresponding probability distribution for the number of pulses, the probabilities $P_n$ can be substituted in Eq. (\ref{Eq:PulseStatVSPhotStat1}).
	The photon-number and pulse-number statistics for these states are compared in Fig. \ref{Fig:SqCS}.
	In this case the effects of dead time and afterpulses are clearly pronounced.
	The corresponding statistics is derived under assumption that the probability of afterpulses $p(|\alpha|^2)$ has approximately the same value $p$ for all values of $\alpha$, where the Wigner function $W(\alpha)$ significantly exceeds zero.   
	
\begin{figure}[ht!]
			\includegraphics[width=\linewidth]{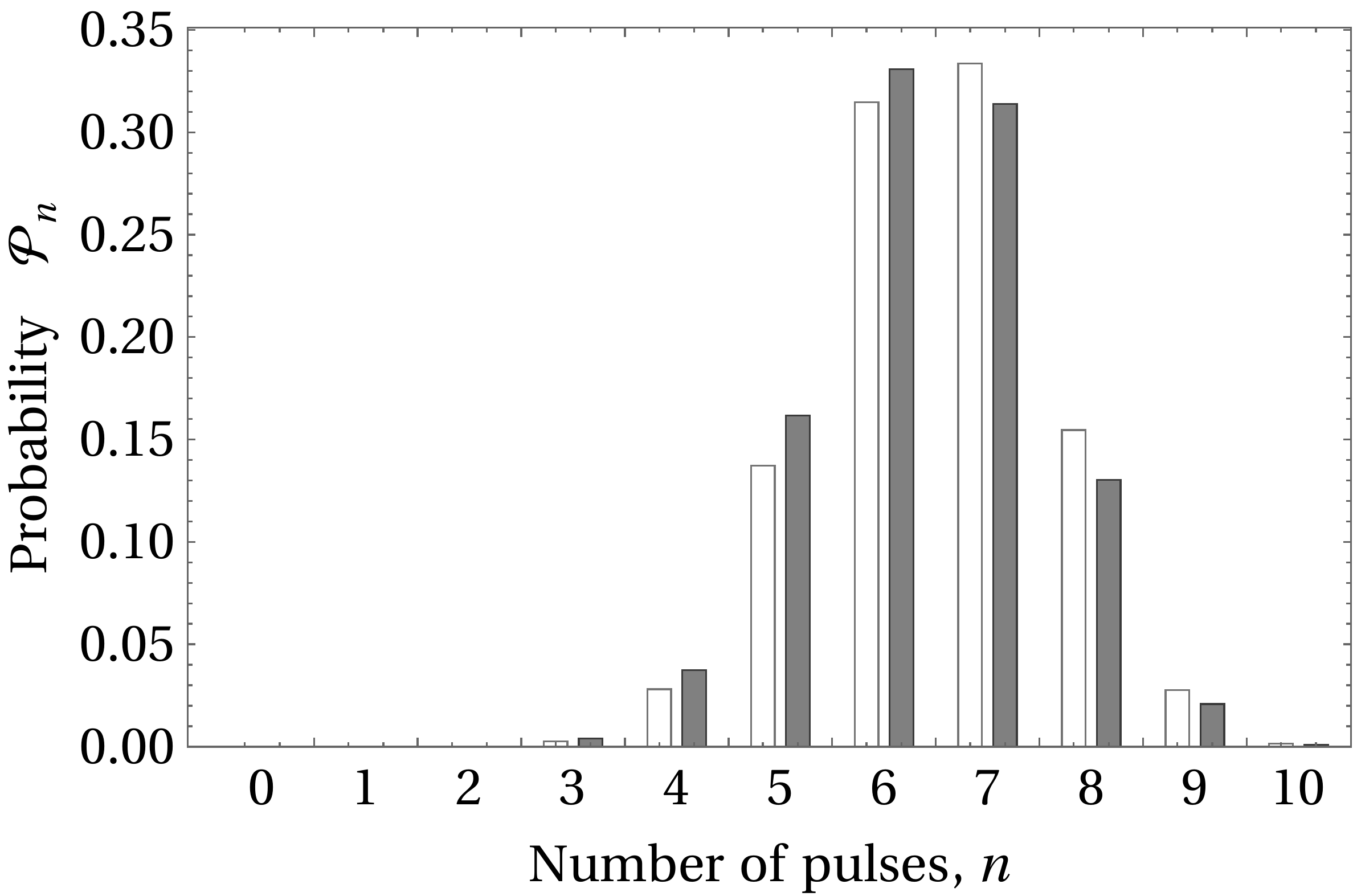}
			\caption{\label{Fig:SqMemP1} The pulse-number distributions, $\mathcal{P}_n$, for the detection with independent MTWs and for the cw detection in the case of attenuated squeezed coherent states given by Eq. (\ref{Eq:PhSqCS}), wherein $r=0.69$, $\eta=0.8$, $\alpha_0=4$, are presented by white and gray bars, respectively.
			Here $\tau_\mathrm{d}=0.09\tau_\mathrm{m}$, $p=0.1$.}
		\end{figure}

	The memory effect from the previous MTWs in the case of cw detection can be considered via employing Eq. (\ref{Eq:PulseStatVSPhotStat2}) together with the approximate expression (\ref{Eq:CondProb_CW}).
	The cases of independent MTWs and the cw detection for a phase-squeezed coherent state are compared in Fig. \ref{Fig:SqMemP1}. 
	In particular, the differences between these distributions are confirmed by the values $\langle n\rangle\approx6.53$, $\langle \Delta n^2\rangle\approx1.22$ and  $\langle n\rangle\approx6.39$, $\langle \Delta n^2\rangle\approx1.24$ for the detection with independent MTWs and the cw detection, respectively. 

\begin{figure}[ht!]
			\includegraphics[width=\linewidth]{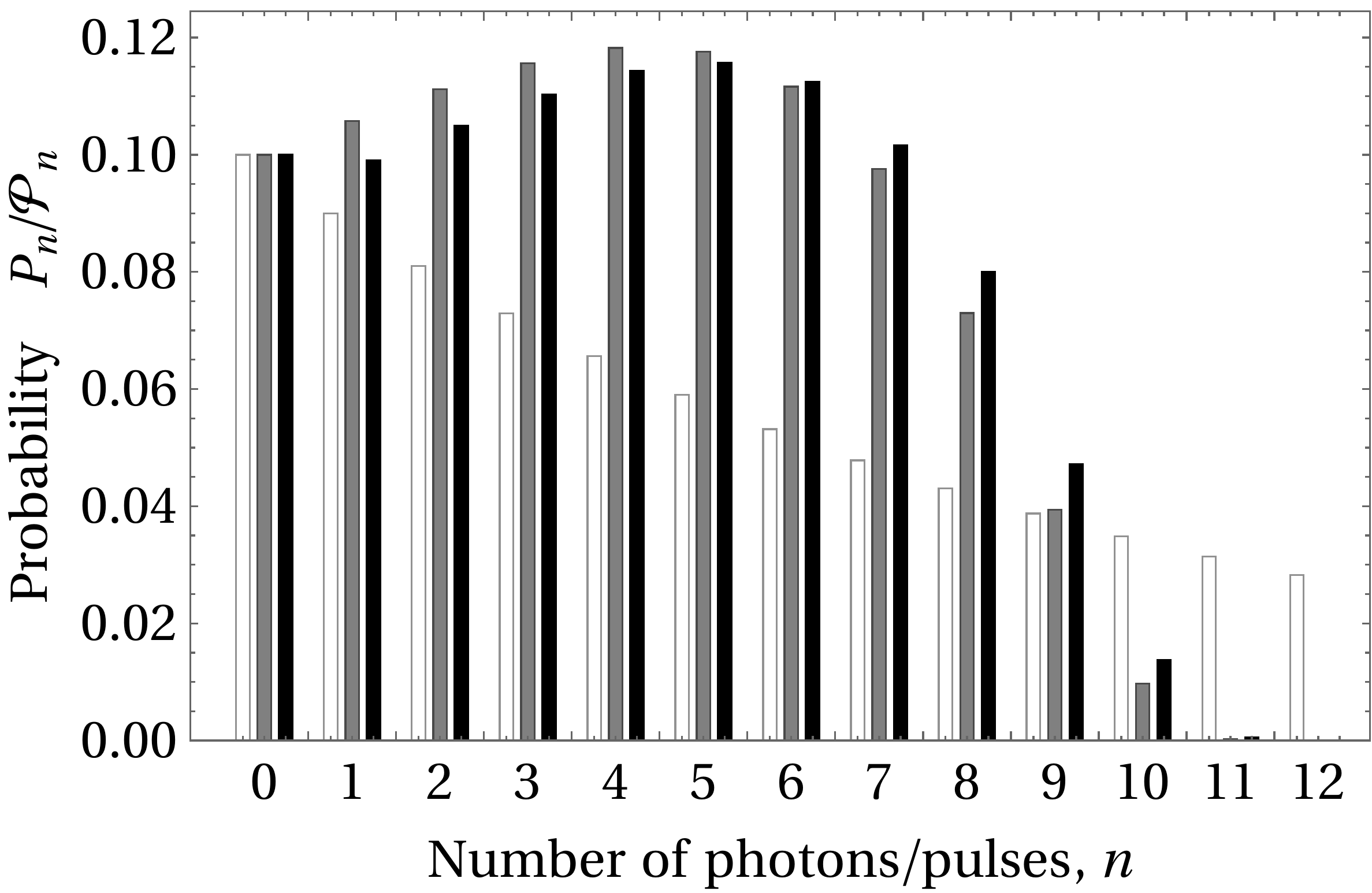}
			\caption{\label{Fig:Th} The photon-number and pulse-number distributions for the thermal state given by Eqs. (\ref{Eq:PhaseLessQS}) and (\ref{Eq:ThSt}), $n_\mathrm{th}=9$, are shown.
			White bars correspond to the photon-number distribution, $P_n$.
			The gray and black bars correspond to the pulse-number distributions, $\mathcal{P}_n$, for the detection with independent MTWs without and with afterpulses, respectively.
			Here $\tau_\mathrm{d}=0.09\tau_\mathrm{m}$, $p=0.07$.
			}
		\end{figure}

\begin{figure}[ht!]
		\includegraphics[width=\linewidth]{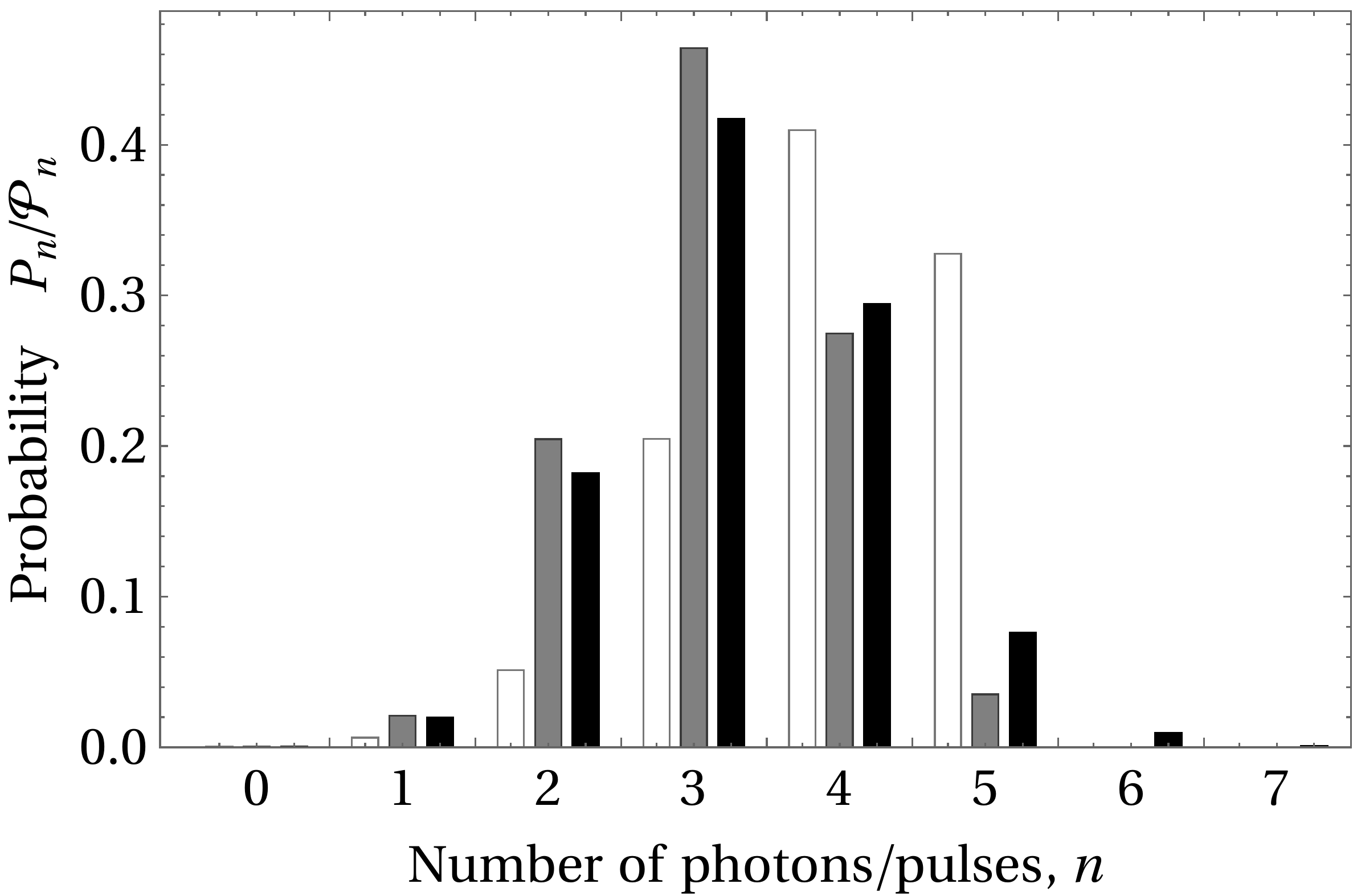}
		\caption{\label{Fig:Fock} The photon-number and pulse-number distributions for the attenuated Fock states given by Eqs. (\ref{Eq:PhaseLessQS}) and (\ref{Eq:FockSt}), $n=5$, $\eta=0.8$, are shown.
		White bars correspond to the photon-number distribution, $P_n$.
		The gray and black bars correspond to the pulse-number distributions, $\mathcal{P}_n$, for the detection with independent MTWs without and with afterpulses, respectively.
		Here $\tau_\mathrm{d}=0.09\tau_\mathrm{m}$, $p=0.07$.
		}
	\end{figure}

	Consider typical examples of phaseless quantum states, which are diagonal in Fock basis,
		\begin{align}
			\hat{\rho}=\sum\limits_{n=0}^{+\infty}P_n\ket{n}\bra{n}.
			\label{Eq:PhaseLessQS}
		\end{align}
	In particular, we are interested in thermal states,
		\begin{align}
			P_n=\frac{1}{1+n_\mathrm{th}}\left(\frac{n_\mathrm{th}}{1+n_\mathrm{th}}\right)^n,
			\label{Eq:ThSt}
		\end{align}
	where $n_\mathrm{th}=\langle \hat{n}\rangle$, and Fock number-states $\ket{l}$ attenuated with the efficiency $\eta$, 
		\begin{align}
			P_n=\binom{l}{n}\eta^n(1-\eta)^{l-n}.
			\label{Eq:FockSt}
		\end{align}	
	As it is clearly seen from Figs. \ref{Fig:Th} and \ref{Fig:Fock}, the dead time and afterpulses may result in pronounced modifications of the corresponding statistics.

%
\section{Expected values of observables}
\label{Sec:Geom}
	
	In many practical applications, it is important to reconstruct the mean values of observables based on the given experimental data.
	In this context, a special case is presented by observables $\hat{A}{=}A(\hat{n})$ commuting with photon-number operator $\hat{n}$.
	In the case of ideal PNR detectors, wherein photocounting statistics coincide with photon statistics $P_n$, such values are evaluated as
		\begin{align}
			\langle \hat{A}\rangle=\sum\limits_{n=0}^{+\infty}A(n)P_n.
		\end{align} 
	However, this trick does not work anymore for realistic detectors.
	
	A way of reconstructing $\langle \hat{A}\rangle$ in the case of imperfect photon-number resolution has been proposed in Ref. \cite{kovalenko2018}. 
	The idea is based on using the POVM as a basis for the space of operators $\hat{A}$. 
	However, for detection techniques limited to a finite number of outcomes, the measurement operators  only span a subspace of the former space of operators, leading to systematic errors in the reconstruction. 
		
	In this section we adapt this technique for the pulse statistics in the scenario of the detection with independent MTWs. 
	In this case, the expected value of the observable $\hat{A}$ is given by
		\begin{align}
			\langle \hat{A}\rangle=\sum\limits_{n=0}^{N_\mathrm{max}-1}A^n\mathcal{P}_n+\mathcal{R}.
			\label{Eq:MeanValueGeom}
		\end{align}	  
	Here, 
		\begin{align}
			N_\mathrm{max}=\left\lceil\frac{\tau_\mathrm{m}}{\tau_\mathrm{d}}\right\rceil=
			\left\{
			\begin{array}{lcl}
				N & \textrm{if} & \tau_\mathrm{m}/\tau_\mathrm{d}\in\mathbb{N}\\
				N+1 & \textrm{if} & \tau_\mathrm{m}/\tau_\mathrm{d}\notin\mathbb{N}
			\end{array}
			\right.,
		\end{align}
	is the maximal number of the detection events in the MTW, $\mathcal{R}$ is a residual term determining the systematic error caused by finiteness of possible number of pulses, $\mathcal{P}_n$ is the pulse-number distribution, and $A^n$ are so-called contravariant coordinates of the observable $\hat{A}$.    
	The latter can be obtained as
		\begin{align}
			A^n=\sum\limits_{m=0}^{N_\mathrm{max}-1}g^{nm}\Tr\left(\hat{A}\hat{\Pi}_m^\mathrm{(r)}\right),
			\label{Eq:ContraVarComp}
		\end{align} 
	where $g^{nm}$ is the so-called contravariant metric tensor, which is inverse to the covariant metric tensor given by
		\begin{align}
			g_{nm}=\Tr\left(\hat{\Pi}_n^\mathrm{(r)}\hat{\Pi}_m^\mathrm{(r)}\right).
		\end{align}
	For the wide class of the Hilbert-Schmidt (HS) operators, the upper bound of the residual term $\mathcal{R}$ can be estimated as the HS mismatch,
		\begin{align}
			\mathcal{R}^2\leq\Tr{\hat{A}^2}-\sum\limits_{n,m=0}^{N_\mathrm{max}-1}g^{nm}\Tr\left(\hat{A}\hat{\Pi}_n^\mathrm{(r)}\right)\Tr\left(\hat{A}\hat{\Pi}_m^\mathrm{(r)}\right).
			\label{Eq:ErrorGeom}
		\end{align}
	This mismatch quantifies to which extent we lose information about the corresponding observable in the considered measurement. 
	As however for operators that do not have an HS norm of one, i.e., $\mathrm{Tr}(\hat{A}^2) \neq 1$, this mismatch scales with the norm, it will make sense to consider the relative HS mismatch, which we get by dividing the right-hand side of Eq. (\ref{Eq:ErrorGeom}) by $\mathrm{Tr}(\hat{A}^2)$.
	
	\subsection{Metric tensor for the detection with independent MTWs}

	For the sake of simplicity, we consider the situation with no afterpulses, no dark counts, and with $\tau_\mathrm{m}/\tau_\mathrm{d}=N=N_\mathrm{max}$.
	Also the detection efficiency will be included in the considered quantum states.
	In this case, the POVM $\hat{\Pi}_n^\mathrm{(r)}$ is given by Eqs. (\ref{Eq:POVM-Ind-Meas-0}) and (\ref{Eq:POVM-Ind-Meas}). 
	This yields the covariant metric tensor in the form
		\begin{align}
			  &g_{nm}= \quad \sum_{k=0}^{n}\sum_{l=0}^{m} G_{k,n;l,m}  - \sum_{k=0}^{n}\sum_{l=0}^{m-1} G_{k,n;l,m-1}\nonumber\\
			&- \sum_{k=0}^{n-1}\sum_{l=0}^{m} G_{k,n-1;l,m} \, + \sum_{k=0}^{n-1}\sum_{l=0}^{m-1} G_{k,n-1;l,m-1},
		\end{align}
	where
		\begin{align}
			G_{k,n;l,m} = \mathrm{Tr}\left(\hat{F}_k\left[\eta_\mathrm{rr}(n)\right]\hat{F}_l\left[\eta_\mathrm{rr}(m)\right]\right).
		\end{align}
	Substituting here the explicit form of $\hat{F}_k\left[\eta_n\right]$, cf. Eq. (\ref{Eq:POVM_PNR}), one gets
		\begin{align}
			&G_{k,n;l,m} = \sum_{i=\mathrm{max}\left\lbrace k,l\right\rbrace}^{k+l} \binom{i}{2i-k-l} \binom{2i-k-l}{i-k}\nonumber \\
			&\times\frac{[\eta_\mathrm{rr}(n)]^{k}[\eta_\mathrm{rr}(m)]^{l}\left[1-\eta_\mathrm{rr}(n)\right]^{i-k}\left[1-\eta_\mathrm{rr}(m)\right]^{i-l}}{\left[ \eta_\mathrm{rr}(n)+\eta_\mathrm{rr}(m)-\eta_\mathrm{rr}(n)\eta_\mathrm{rr}(m) \right]^{i+1}}.
		\end{align}
	In these equations the adjusting efficiency $\eta_\mathrm{rr}(n)$ is given by Eq. (\ref{Eq:Eta_rr}).
	
	As it has been discussed in Ref. \cite{kovalenko2018}, the covariant metric tensor of the finite-dimensional measurements necessarily should have at least one singular element.
	In the considered case this is $g_{NN}=\infty$.
	This property implies that the corresponding contravariant metric tensor satisfies the conditions
		\begin{align}
			g^{Nm}=g^{nN}=g^{NN}=0.
		\end{align}
	Also the relation
		\begin{align} 
			\sum\limits_{k=0}^{N-1}g^{mk}g_{kn}=\delta_{n}^{m}
			\label{Eq:InversTensor}
		\end{align}
	holds true that justifies the upper bound of the sums in Eqs. (\ref{Eq:MeanValueGeom}), (\ref{Eq:ContraVarComp}), and (\ref{Eq:ErrorGeom}).
	Another important feature of the metric tensor is the fact that $g_{00}=g^{00}=1$ and $g_{0n}=g^{0n}=0$ for all $n\neq 0$.

	\subsection{Application to unbalanced homodyne detection}

\begin{figure}[ht!]
			\includegraphics[width=0.9\linewidth]{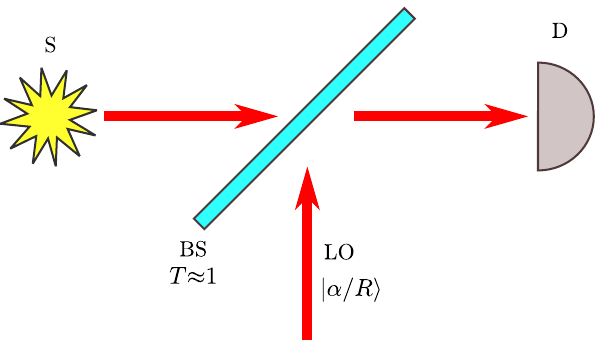}
			\caption{\label{Fig:UHD} The scheme of unbalanced homodyne detection.
			The signal mode (S) is interfered at the beam splitter (BS) with the transmission coefficient $T$ close to unity with the field of a local oscillator (LO) in the coherent state $\ket{\alpha/R}$, where $R$ is the reflection coefficient of the beam splitter.
			The resulting state displaced in the phase with the coherent amplitude $\alpha$ is analyzed at the detector~D.}
		\end{figure}

	As an example, demonstrating the developed techniques, we consider their application to the quantum-state reconstruction scheme based on unbalanced homodyne detection, see Refs. \cite{wallentowitz96,mancini1997}.
	The idea consists in the fact that the value of the Cahill-Glauber $s$-parametrized quasiprobability distribution $P(\alpha;s)$ \cite{cahill69,cahill69a} in the phase-space point $\alpha$ can be interpreted as the expectation value of the operator
		\begin{align}
			\hat{P}(\alpha;s)=\frac{2}{\pi(1-s)}:\exp\left[-\frac{2}{1-s}\hat{n}(\alpha)\right]:,
			\label{Eq:PS_Kern}
		\end{align}  
	where $\hat{n}(\alpha)=(\hat{a}^\dag-\alpha^\ast)(\hat{a}-\alpha)$ is the displaced photon-number operator.
	The whole measurement procedure includes displacement of quantum states in phase space and the detector, analyzing the resulting light mode, see Fig.~\ref{Fig:UHD}.

	The covariant components of the operator (\ref{Eq:PS_Kern}) can be directly calculated as
		\begin{align}
			&\Tr\left[\hat{P}(0;s)\hat{\Pi}_n^\mathrm{(r)}\right]=\frac{2}{\pi(1-s)}\nonumber\\
			&\times\left[
			\sum\limits_{k=0}^{n}\frac{[-\eta_\mathrm{rr}(n)]^k(s+1)^k(1-s)}{[2-(1+s)\eta_\mathrm{rr}(n)]^{k+1}}
			\right.\nonumber\\
			&-\left.
			\sum\limits_{k=0}^{n-1}\frac{[-\eta_\mathrm{rr}(n-1)]^k(s+1)^k(1-s)}{[2-(1+s)\eta_\mathrm{rr}(n-1)]^{k+1}}
			\right].
		\end{align}
	This relation is invariant with respect to simultaneous displacements of the coherent amplitude $\alpha$ in the operators $\hat{P}(0;s)$ and $\hat{\Pi}_n^\mathrm{(r)}$.  
	Let $\mathcal{P}_n(\alpha)$ be the pulse-number distribution for the considered state displaced to the phase-space point $\alpha$.
	Then, Eq. (\ref{Eq:MeanValueGeom}) yields the estimation of the Cahill-Glauber $s$-parametrized distribution
		 \begin{align}
		 	P(\alpha;s)\approx \sum\limits_{n,m=0}^{N-1}g^{nm}\Tr\left[\hat{P}(0;s)\hat{\Pi}_m^\mathrm{(r)}\right]\mathcal{P}_n(\alpha).
		 	\label{Eq:PS_Reconstr}
		 \end{align}
	The systematic error of this procedure can be estimated as the HS mismatch according to Eq. (\ref{Eq:ErrorGeom}).
	
	Let us illustrate the applicability of the method by simulating the reconstruction of the phase-space distribution of an attenuated squeezed vacuum state, given by the Wigner function (\ref{Eq:PhSqCS}) with $\alpha^\prime_0=0$.
	We simulate $M$ random values of pulse numbers $n_i$ related to the probability distributions $\mathcal{P}_n(\alpha)$ for some values of $\alpha$.
	After that we estimate the approximated probability distribution as
		\begin{align}
			\mathcal{P}_n(\alpha)\approx\frac{1}{M}\sum_{i=1}^{M}\delta_{nn_i}.
		\end{align}
	This probability distribution is used in Eq. (\ref{Eq:PS_Reconstr}) for reconstruction of the $s$-parametrized phase-space distribution.    
	
\begin{figure}
		\includegraphics[width=\linewidth]{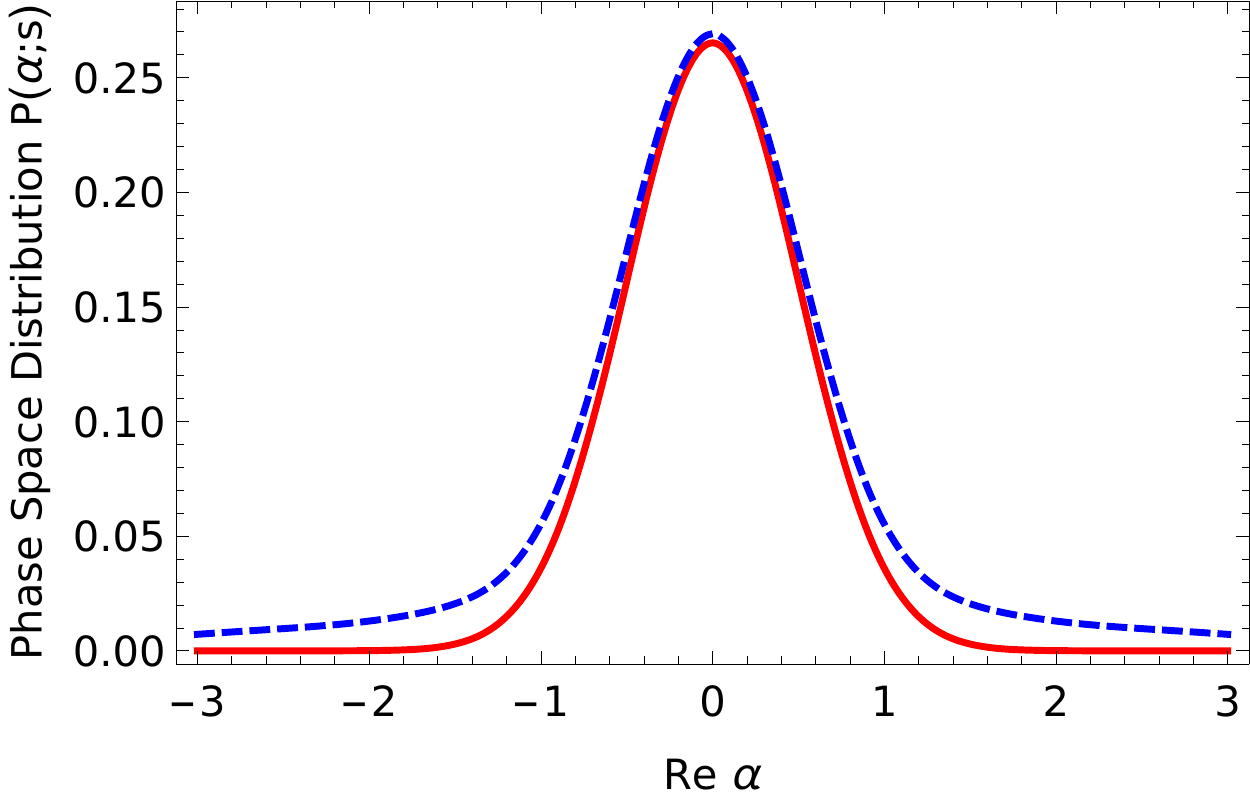}
		\caption{\label{Fig:Rec-UHD-P} Reconstruction of the phase-space distribution $P(\alpha,s)$ with $s=-0.8$ on the real axis of a squeezed vacuum state with squeezing parameter $r=0.8$ for detection with a maximal number of counting events per interval of $N=8$. The reconstructed distribution (dashed line) is compared with the theoretical distribution of the state (solid  line).}
	\end{figure}

\begin{figure}
		\includegraphics[width=\linewidth]{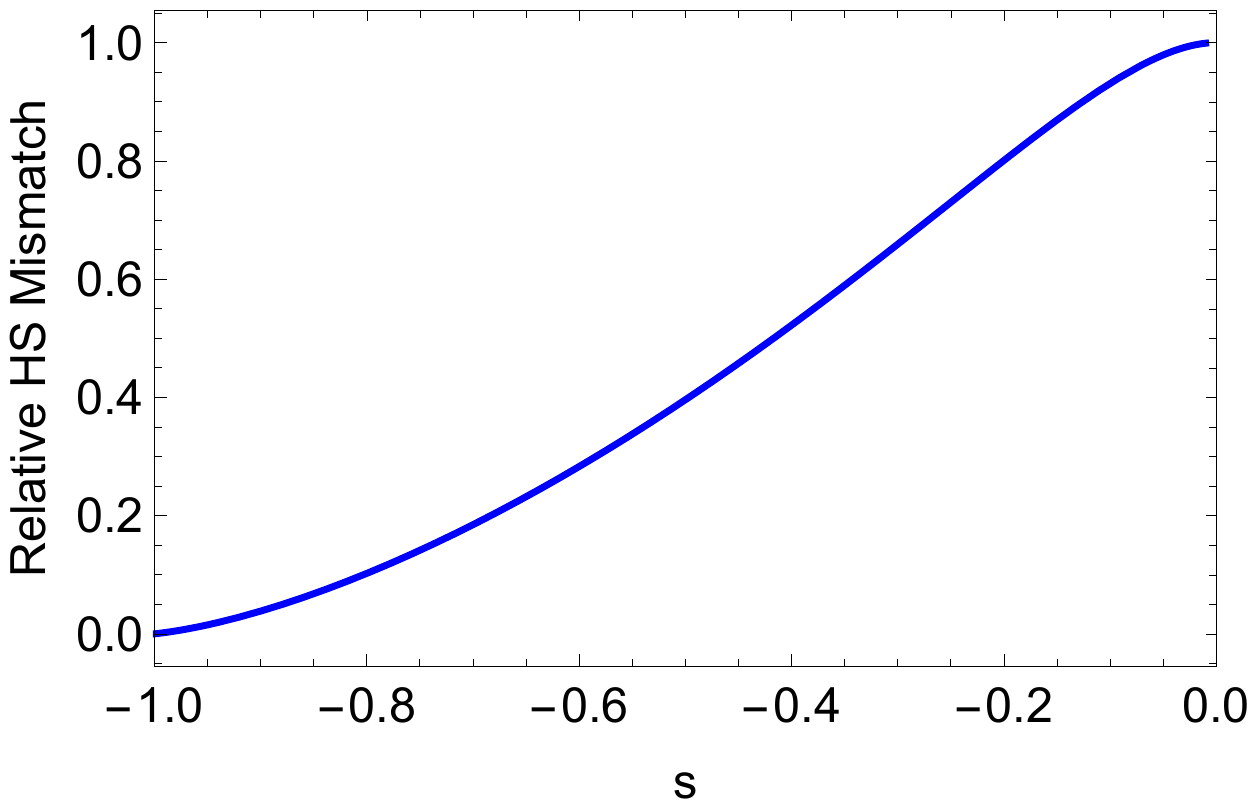}
		\caption{\label{Fig:HS-UHD-relative} The relative HS mismatch for the operator $\hat{P}(\alpha,s)$ is plotted for the values of $s$ ranging from -1 to 0 and a maximal number of counting events per MTW for $N=8$. }
	\end{figure}

	An example of such a reconstruction compared with the original theoretical phase-space distribution is shown in Fig. \ref{Fig:Rec-UHD-P}.
	It is clearly seen that the procedure gives a satisfactory result near the phase-space origin and a larger error for other values of $\alpha$.
	The relative HS mismatch for such a procedure is increasing with increasing the parameter $s$, as shown in Fig. \ref{Fig:HS-UHD-relative}.  
	
\section{Statistics of the time between pulses}	
\label{Sec:Time}	
	
	In this section we consider statistics of the time between two subsequent pulses of photocurrent, $t_\mathrm{b}$.
	The probability distribution of this random variable can be straightforwardly reconstructed in the experiment. 
	This distribution explicitly depends on characteristics of the detection process, such as the dead time $\tau_\mathrm{d}$ and the probability of afterpulses, $p$. 
	Hence, we obtain a possibility to estimate the values of these quantities and get other characteristics of the radiation field and the detection process.
	
	Let us consider the electromagnetic radiation, corresponding to the coherent state $\ket{\alpha_0}$ in the MTW of duration $\tau_\mathrm{m}$.
	For the sake of simplicity, we consider that detection losses are included in the signal attenuation such that the coherent amplitude $\alpha_0$ is rescaled with the detection efficiency. 
	In this case, the photon flux $\lambda=|\alpha_0|^2/\tau_\mathrm{m}$ is the mean number of photons per a time unit.
	The inverse quantity $\tau_\mathrm{phot}=\lambda^{-1}$ defines the mean time passing between registrations of two photons with the ideal PNR detector.
	
	If the first pulse in a couple of two neighboring pulses has been registered, then the probability to register the second one is zero inside the dead-time interval.
	Outside this interval, the infinitesimal probability to register the second pulse during the time interval $dt$ is obtained similar to Eq.~(\ref{Eq:InfProb}) and given by
		\begin{align}
			\frac{dt}{\tau_\mathrm{phot}}\exp\left(-\frac{dt}{\tau_\mathrm{phot}}\right)\rightarrow\frac{dt}{\tau_\mathrm{phot}}.
		\end{align}
	Multiplying this quantity by the probability of no-photon registration in the time interval $[\tau_\mathrm{d},t_\mathrm{b}]$ given by $\exp[-\tau_\mathrm{phot}^{-1}(t_\mathrm{b}-\tau_\mathrm{d})]$, we arrive at the probability density of the time between subsequent pulses,
		\begin{align}\label{Eq:TimePhot}
			P_\mathrm{phot}(t_\mathrm{b})=\frac{1}{\tau_\mathrm{phot}}\exp\left[-\frac{t_\mathrm{b}-\tau_\mathrm{d}}{\tau_\mathrm{phot}}\right],
		\end{align}
	defined outside the dead-time interval $[0,\tau_\mathrm{d}]$.	
	
	Equation (\ref{Eq:TimePhot}), however, does not consider the possibility of a registration of an afterpulse at the time moment $\tau_\mathrm{d}$.
	The corresponding probability density in this case is given by the Dirac delta-distribution,
		\begin{align}\label{Eq:TimeAfter1}
			P_\mathrm{after}(t_\mathrm{b})=\delta(t_\mathrm{b}-\tau_\mathrm{d}).
		\end{align}
	Considering the probabilities of registering and nonregistering an afterpulse to be equal $p$ and $(1-p)$, respectively, we arrive at the probability distribution of the time between consecutive pulses, which includes also the presence of afterpulses,
		\begin{align}\label{Eq:TimePDF1}
			P(t_\mathrm{b})=\theta(t_\mathrm{b}&-\tau_\mathrm{d})\nonumber\\
			&\times\left[(1-p)P_\mathrm{phot}(t_\mathrm{b})+pP_\mathrm{after}(t_\mathrm{b})\right].
		\end{align}  
	Here $\theta(t_\mathrm{b}-\tau_\mathrm{d})$ is the Heaviside step-function corresponding to the zero probability of registering photons during the dead-time interval.

	For comparing experimental data with the theoretical model, it is convenient to use the cumulative probability distribution, $\mathcal{F}(t_\mathrm{b})=\int_{0}^{t_\mathrm{b}}dt^\prime P(t^\prime)$, which reads
		\begin{align}\label{Eq:TimeCDF1}
			\mathcal{F}(t_\mathrm{b})=(1-p)\mathcal{F}_\mathrm{phot}(t_\mathrm{b})+p\mathcal{F}_\mathrm{after}(t_\mathrm{b}).
		\end{align}
	In the considered scenario
		\begin{align}\label{Eq:TimePhotComm}
			\mathcal{F}_\mathrm{phot}(t_\mathrm{b})=\theta(t_\mathrm{b}-\tau_\mathrm{d})\left\{1-\exp\left[-\frac{t_\mathrm{b}-\tau_\mathrm{d}}{\tau_\mathrm{phot}}\right]\right\},
		\end{align} 
		\begin{align}\label{Eq:TimeAfterComm1}
			\mathcal{F}_\mathrm{after}(t_\mathrm{b})=\theta(t_\mathrm{b}-\tau_\mathrm{d}).
		\end{align}   
	Since the distribution function explicitly depends on the parameters $p$ and $\tau_\mathrm{d}$, they can be estimated from the experimental data.
	
	Let us consider the mean time between pulses,
		\begin{align}
			\left\langle t_\mathrm{b} \right\rangle=\int\limits_{0}^{+\infty} dt_\mathrm{b}t_\mathrm{b}P(t_\mathrm{b})=(1-p)(\tau_\mathrm{phot}+\tau_\mathrm{d}).
		\end{align}
	This relation can be used for an estimation of the photon flux given $p$ and $\tau_\mathrm{d}$ as
		\begin{align}
			\lambda=\tau_\mathrm{phot}^{-1}=\frac{1-p}{\left\langle t_\mathrm{b} \right\rangle-\tau_\mathrm{d}(1-p)}.
		\end{align}	
	It is worth noting that in the absence of afterpulses $\tau_\mathrm{phot}=\left\langle t_\mathrm{b} \right\rangle-\tau_\mathrm{d}$.
	
	Fitting experimental data related to the time between pulses may require small modifications of the considered model. 
	For example, after ending the dead-time interval, the ability of detectors to register the next photon is not recovered immediately. 
	Similar to superconducting nanowire single-photon detectors (SNSPDs), this recovering can be described by the time-dependent detection efficiency \cite{Uzunova2022},
		\begin{align}\label{Eq:TDE}
			\eta_\mathrm{r}(t)=1-\exp\left(-\frac{t}{\tau_\mathrm{r}}\right).
		\end{align}
	Unlike the case of the SNSPDs, the recovering time $\tau_\mathrm{r}$ for the avalanche photodiods is significantly less than the dead time $\tau_\mathrm{d}$. 
	This recovering process does not effect the photocounting statistics.
	However, it may still be taken into account for fitting the statistics of time between pulses. 
	
	Similarly to the consideration in Ref.~\cite{Uzunova2022}, the probability distribution $P_\mathrm{phot}(t_\mathrm{b})$ is modified as
		\begin{align}\label{Eq:TimePhotM}
			P_\mathrm{phot}(t_\mathrm{b})=\frac{\eta_\mathrm{r}(t_\mathrm{b}-\tau_\mathrm{d})}{\tau_\mathrm{phot}}\exp\left[-\frac{\Xi(t_\mathrm{b})}{\tau_\mathrm{phot}}\right],
		\end{align}
	where	
		\begin{align}
			\Xi(t_\mathrm{b})=\int\limits_{\tau_\mathrm{d}}^{t_\mathrm{b}}dt\eta_\mathrm{r}(t-\tau_\mathrm{d}).
		\end{align}
	The corresponding cumulative probability distribution reads
		\begin{align}\label{Eq:TimePhotCommM}
			\mathcal{F}_\mathrm{phot}(t_\mathrm{b})=\theta(t_\mathrm{b}-\tau_\mathrm{d})
			\left\{1-\exp\left[-\frac{\Xi(t_\mathrm{b})}{\tau_\mathrm{phot}}\right]\right\}.
		\end{align}
	For the model of time-dependent efficiency given by Eq.~(\ref{Eq:TDE}), one gets
		\begin{align}
			\Xi(t_\mathrm{b})=t_\mathrm{b}-\tau_\mathrm{d}-\tau_\mathrm{r}
			\left[1-\exp\left(-\frac{t_\mathrm{b}-\tau_\mathrm{d}}{\tau_\mathrm{r}}\right)\right].
		\end{align}
	If the recovering time vanishes, $\tau_\mathrm{r}=0$, Eqs.~(\ref{Eq:TimePhotM}) and (\ref{Eq:TimePhotCommM}) becomes Eqs.~(\ref{Eq:TimePhot}) and (\ref{Eq:TimePhotComm}), respectively. 
	
	Another modification to the model given by Eqs.~(\ref{Eq:TimePDF1}) and (\ref{Eq:TimeCDF1}) is related to the fact that an afterpulse may not appear immediately after ending the dead-time interval.
	This means that Eq.~(\ref{Eq:TimeAfter1}) should be modified from the form with the delta distribution to the form similar to Eq.~(\ref{Eq:TimePhot}) by replacing the time $\tau_\mathrm{phot}$ related to the photon flux by another time $\tau_\mathrm{after}$ related to the afterpulses. 
	Moreover, the ability of the detector to register the afterpulse is also recovered smoothly with the recovering time $\tau_\mathrm{r}$ similarly to the same property for the photon-related pulses.
	This implies that the probability distribution $P_\mathrm{after}(t_\mathrm{b})$ is now given by
		\begin{align}\label{Eq:TimeAfterM}
			P_\mathrm{after}(t_\mathrm{b})=\frac{\eta_\mathrm{r}(t_\mathrm{b}-\tau_\mathrm{d})}{\tau_\mathrm{after}}\exp\left[-\frac{\Xi(t_\mathrm{b})}{\tau_\mathrm{after}}\right].
		\end{align}  
	The corresponding cumulative probability distribution reads
		\begin{align}\label{Eq:TimeAfterCommM}
			\mathcal{F}_\mathrm{after}(t_\mathrm{b})=\theta(t_\mathrm{b}-\tau_\mathrm{d})
			\left\{1-\exp\left[-\frac{\Xi(t_\mathrm{b})}{\tau_\mathrm{after}}\right]\right\}.
		\end{align}
	For $\tau_\mathrm{after}=\tau_\mathrm{r}=0$ these equations take the form of Eqs.~(\ref{Eq:TimeAfter1}) and (\ref{Eq:TimeAfterComm1}), respectively.


\section{Experimental verification}
\label{Sec:Experiment}

	In this section we report about an experimental verification of the theory presented in previous sections.
	In this context, we perform four stages of the experimental-data processing.
	First, we reconstruct the empirical probability distribution for the time between pulses,
		\begin{align}\label{Eq:EmpiricPD}
			\mathcal{F}(t_\mathrm{b})=\frac{1}{K}\sum\limits_{k=1}^{K}\theta(t_\mathrm{b}-t_\mathrm{b}^{(k)}),
		\end{align}
	where $t_\mathrm{b}^{(k)}$ are $K$ measured time intervals between pulses. 
	We estimate values of the photon flux $\lambda=\tau_\mathrm{phot}^{-1}$, the dead time $\tau_\mathrm{d}$, the probability $p$ of afterpulses, the afterpulse time constant $\tau_\mathrm{after}$, and the recovering time $\tau_\mathrm{r}$ via fitting the function (\ref{Eq:EmpiricPD}) with Eqs.~(\ref{Eq:TimeCDF1}), (\ref{Eq:TimePhotCommM}), and (\ref{Eq:TimeAfterCommM}). 
	Second, we study the dependence of $\tau_\mathrm{d}$, $p$, $\tau_\mathrm{after}$, and $\tau_\mathrm{r}$ on the photon flux $\lambda$, i.e., on the mean photon-number $|\alpha|^2$ in order to conclude either this dependence should be included in the photocounting formula.
	Third, we directly check the assumption of uniform distribution (\ref{Eq:RhoUniform}) for $\varrho_l(\tau_2;\boldsymbol{\alpha}^{l-1})$.
	Fourth, we reconstruct the photocounting statistics and compare it with the derived photocounting formulas.
	For this purpose, we use the values of $|\alpha|^2$, $\tau_\mathrm{d}$, and $p$ estimated at the first stage.
	
	For detecting photons we used a silicon avalanche photodiode module ($\tau$-\textsf{SPAD} by \textsf{PicoQuant}).
	Coherent states of light were generated with continuous coherent illumination from an Nd:YAG laser (\textsf{Mephisto S} by \textsf{Innolight}) at a wavelength of \SI{1064}{\nano\metre}. 
	The output signal was continuously sampled with an analog to digital converter at a sampling rate of \SI{200}{\mega\siemens\per\second} and recorded for later processing.

\begin{figure}[ht!]
		\includegraphics[width = \linewidth]{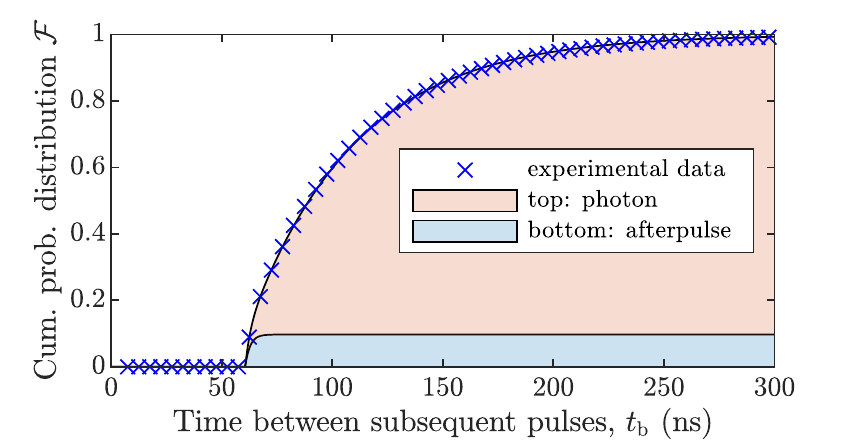}
		\caption{\label{Fig:CPD} Experimentally-estimated empirical probability distribution [cf. Eq.~(\ref{Eq:EmpiricPD})], fitted theoretical model [cf. Eqs.~(\ref{Eq:TimeCDF1}), (\ref{Eq:TimePhotCommM}), and (\ref{Eq:TimeAfterCommM})], and the explicit theoretical contribution from afterpulses, $p\mathcal{F}_\mathrm{after}(t_\mathrm{b})$.
		In this case the values of parameters are estimated as $\tau_\mathrm{phot}=48.77$ns, $\tau_\mathrm{d}=60.54$ns, $p=0.0972$, $\tau_\mathrm{after}=2.20$ns, and $\tau_\mathrm{r}=0.12$ns.}
\end{figure}

	An example of the fit results for the cumulative probability distribution of the time between pulses with the model given by Eqs.~(\ref{Eq:TimeCDF1}), (\ref{Eq:TimePhotCommM}), and (\ref{Eq:TimeAfterCommM}) is shown in Fig.~\ref{Fig:CPD}.
	The model demonstrates a good agreement with the experimental data.
	The related values of the parameters are given in the caption to this figure.

\begin{figure}[ht!]
		\includegraphics[width = \linewidth]{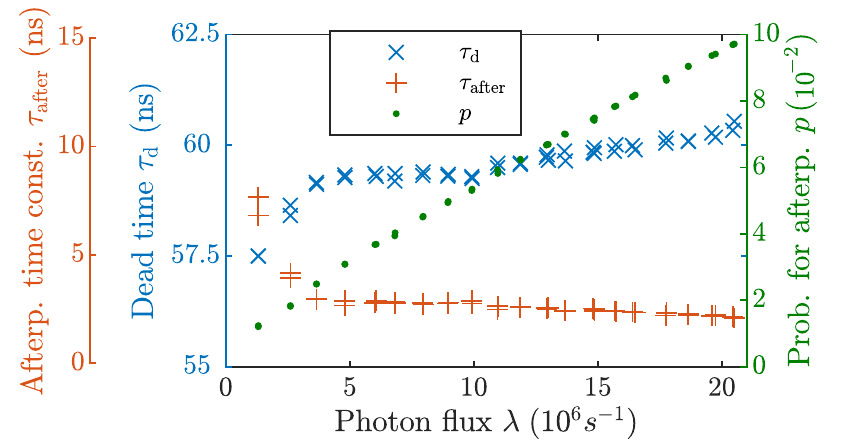}
		\caption{\label{Fig:Fitparams} Detector parameters from fitting the model of Eqs.~(\ref{Eq:TimeCDF1}), (\ref{Eq:TimePhotCommM}), and (\ref{Eq:TimeAfterCommM}) to experimental data. 
		The photon flux is also derived from the fit results such that the detection efficiency is included in the signal attenuation.
	 	}
		\end{figure}

	We find a dependence of the parameters on the incident number of photons which we varied from roughly 1 to 20 photons per \si{\micro\second} corresponding to roughly 0.05 to 1.2 photons per dead time interval. 
	The fitted parameters are shown in Fig. \ref{Fig:Fitparams}. 
	The dead time $\tau_\mathrm{d}$ and the afterpulse time constant $\tau_\mathrm{after}$ exhibit largely negligible variations. 
	However, the probability for afterpulses $p$ depends significantly (apparently linear) on the photon flux,
		\begin{align}
			p\left(|\alpha|^2\right)=p_0+\frac{r}{\tau_\mathrm{m}}|\alpha|^2,
		\end{align}
	where $p_0$ and $r$ are constants characterizing the detection process.
	For the considered case they are estimated as $p_0=0.0093$ and $r=4.4\times10^{-9}$s.
	We hypothesize that this correlation is linked to the detector thermal heating, which increases linearly with the incident photon flux.
	This dependence may be omitted if $p_0\tau_\mathrm{m}/r+\langle \hat{n}\rangle\gg {(|\langle:\Delta \hat{n}^2:\rangle|)^{1/2}}$.   
	Otherwise, it should be explicitly included in the derived photoconting formulas.  
	
	As it is discussed in Sec.~\ref{Sec:GenTheorMod}, an important concept in theoretical model of the continuous-wave detection comprises the influence of the detected photons in the previous MTWs on the pulse-number distribution in the actual one. 
	The dead-time interval of a pulse at the very end of the $(l-1)$th MTW can leak into the $l$th MTW. 
	This casts the detector in its dead state at the beginning of  the $l$th MTW during the time $\tau_2$.
	The probability distribution of this time $\varrho_l(\tau_2;\boldsymbol{\alpha}^{l-1})$ is an integral part of the multiwindow POVM (\ref{Eq:POVM_Gen}).
	Since the process is ergodic, we can omit the dependence on $l$ and consider this probability distribution as $\varrho(\tau_2;\boldsymbol{\alpha})$, where $\boldsymbol{\alpha}$ includes coherent amplitudes from all previous MTWs.
	
	The approximation $\varrho(\tau_2;\boldsymbol{\alpha})$ by the uniform distribution (\ref{Eq:RhoUniform}) enables one to find an analytical form of the multiwindow POVM, cf. Eq.~(\ref{Eq:POVM_Appr}).
	As it follows from the normalization condition (\ref{Eq:Normalization}), the cumulative probability distribution in this case reads
		\begin{align}\label{Eq:CPDUniform}
			\mathcal{F}(\tau_2;\boldsymbol{\alpha})=\mathcal{Q}\left(\boldsymbol{\alpha}\right)+\frac{1-\mathcal{Q}\left(\boldsymbol{\alpha}\right)}{\tau_\mathrm{d}}\tau_2,
		\end{align}      
	where $\mathcal{Q}\left(\boldsymbol{\alpha}\right)$	is the probability that no dead-time interval exceeds the MTW.
	This number explicitly depends on the parameters $\tau_\mathrm{d}$ and $p$, cf. Eq.~(\ref{Eq:ProbabilityQ}) and the definitions of its constituents (\ref{Eq:Al}), (\ref{Eq:Bl}), and (\ref{Eq:Cl}).
	These parameters can be taken from the estimation procedure described above.
	
\begin{figure}[ht!]
			\includegraphics[width=\linewidth]{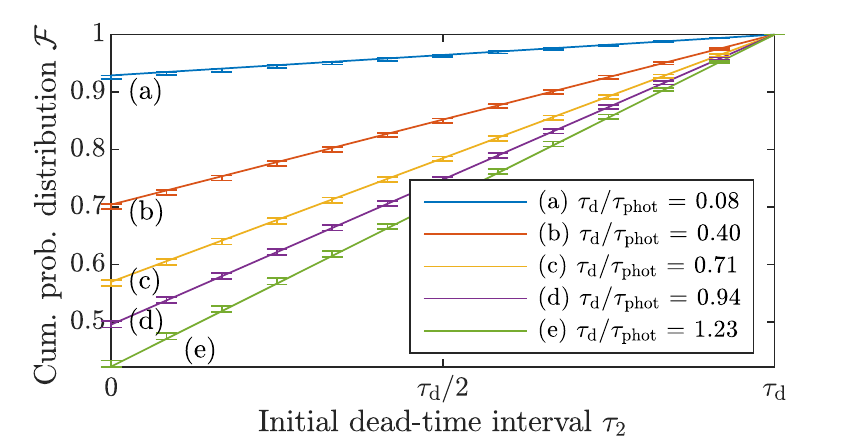}
			\caption{\label{Fig:CDF} Cumulative probability distribution for the initial dead-time interval, $\mathcal{F}(\tau_2;\boldsymbol{\alpha})$, is given for theoretical uniform distribution (\ref{Eq:CPDUniform}) (solid lines), and the estimated from experimental data empiric probability distribution (\ref{Eq:EmpiricPDTau2}) (errorbars). 
			}
		\end{figure}

	We compare experimentally the empiric probability distribution,
		\begin{align}\label{Eq:EmpiricPDTau2}
			\mathcal{F}(\tau_2)=\frac{1}{L}\sum\limits_{l=1}^{L}\theta(\tau_2-\tau_2^{(l)}),
		\end{align}
	where $\tau_2^{(l)}$ are measured times $\tau_2$ from $L$ MTWs, with the theoretical approximation (\ref{Eq:CPDUniform}). 	
	The result for different values of the photon flux is given in Fig.~\ref{Fig:CDF}.
	One can see that the approximation by the uniform distribution demonstrates a good agreement with the experimental data.    	
		
		\begin{figure}[h!!!]
			\includegraphics[width = \linewidth]{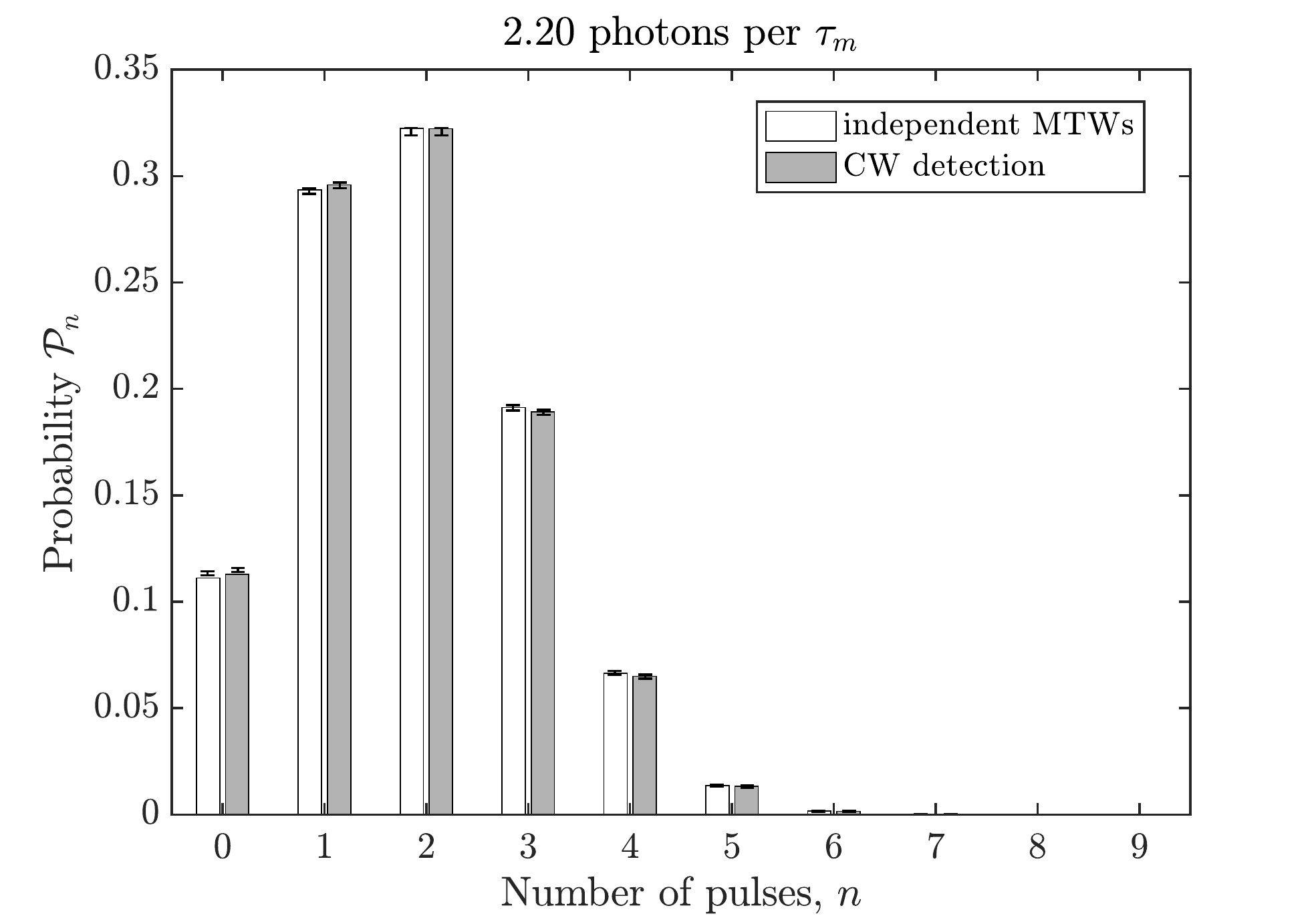}\\
			\includegraphics[width = \linewidth]{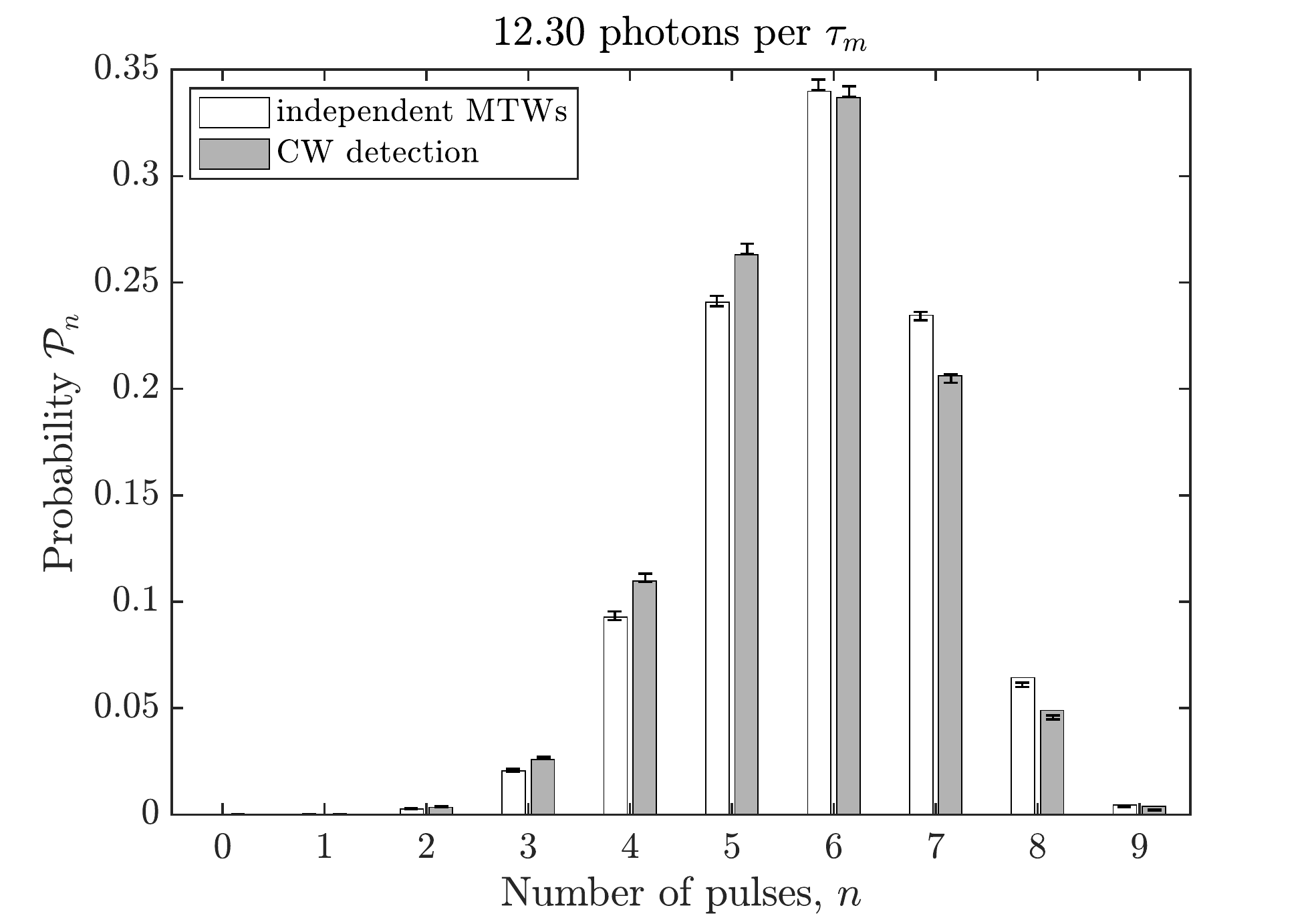}
			\caption{\label{Fig:fitparams} Pulse number statistics for coherent states is shown.
			Bars are related to the theory with independently-estimated parameters. Markers and errorbars are obtained from the experimental data. 
			White and gray bars corresponds to the scenarios with independent MTWs and the cw detection, respectively.
			The parameter values are $\tau_\mathrm{d}=59.12$ ns, $p=0.025$ and  $\tau_\mathrm{d}=60.54$ ns, $p=0.097$ for the upper and lower histograms, respectively.
		}
		\end{figure}

	Finally, we compare the pulse-number statistics we obtained experimentally with coherent illumination of varying strength against the theoretical model of Secs.~\ref{Sec:Recov} and \ref{Sec:CWD}. 
	At this point we want to stress, that the comparison has a quasi ab-initio character. 
	We do not fit the experimental click statistics to the theoretical model. 
	We rather put in the fitted parameters $\tau_\mathrm{d}$, $p$, and $|\alpha|^2=\tau_\mathrm{m}/\tau_\mathrm{phot}$ from the previously described study of the statistics of the time between subsequent pulses into the theoretical model.  
	Two examples from the low end and the high end of the range of photon fluxes are shown in Fig.~\ref{Fig:fitparams}. 
 
    We conclude, that we observed no statistically significant deviation between theory and experiment, and that the theoretical model is a bonafide description of our experimental observation.
    We use the standard 95\% confidence interval for the error bars.
    A small discrepancy between the experimental data and theory in Fig.~\ref{Fig:fitparams} (the case with 12.30 photons per $\tau_\mathrm{m}$) can be caused by imperfections not included in our model, see Refs.~\cite{straka20,Hlousek2023}, by small deviations (near 2\%) of the coherent amplitude over the data collection time, or by imperfect estimation of the model parameters through the fitting presented in Fig.~\ref{Fig:CPD}.
	At this stage we conclude, that the theoretical model describes experimentally obtained click statistics with continuous illumination appropriately.


\section{Conclusions}
\label{Sec:Conclusions}

	To conclude, we have derived the photocounting formula for the detection technique consisting in counting the number of photocurrent pulses inside a MTW and verified its validity in experiments.
	In particular, this formula includes effects of the detector dead time, afterpulses, and the memory effect from the previous MTWs.
	The latter implies that the direct product of density operators from previous MTWs is included in the formula.
	This results in a nonlinear dependence of the photocurrent-pulse statistics from the quantum state of an electromagnetic-field mode. 
	In the most general case, constituents of the derived photocounting formula can be found as a solution to a system of linear recursive equations.
	We have also found an approximation for this solution in the case of small number of photons per dead-time interval.
	It has been shown that only a few previous MTWs affect on the photocounting statistics.
	Hence, the photocounting process is ergodic.
	This fact theoretically justifies the widely-used experimental practice, wherein the photocounting statistics is assumed to be equal for all MTWs.    
	
	Fock-state representation of the photocounting formula determines the relation between photocurrent pulse and photon statistics.
	This expression is nonlinear in the case of cw detection that describes the memory effect from the previous MTWs.
	However, it has an ordinary linear character for the measurements with independent MTWs, assuming that each MTW is followed by a time interval with darkening detector input.
	
	The scheme with independent MTWs is described by the photocounting formula, having the standard linear dependence on the density operator.
	This gives a possibility to use it in a variety of procedures involving data from detectors with imperfect resolution of photonumbers.
	Particularly, we have demonstrated with numerical simulations that the obtained data can be used for quantum-state reconstruction with unbalanced homodyne detection.   
	
	Our theory as well as the used approximation of the uniform distribution for the initial dead-time interval have been directly checked experimentally.
	We have performed an independent estimation of parameters, describing the detection process from analysis of statistics of time between pulses.
	The values of these parameters have been substituted in our theoretical formulas.
	The latter has shown to be in good agreement with the directly-sampled experimental data.
	We believe that the derived photocounting formula can be implemented in different applications, involving the considered type of detection.
	
	A.A.S. acknowledges support from Department of Physics and Astronomy of the NAS of Ukraine through the Project 0120U101347 and from Simons Foundations through Presidential Discretionary-Ukraine Support grant 1030283.
	Ch.B. thanks the Deutscher Akademisher Austauschdienst (DAAD) for supporting this work through a PROMOS scholarship. B.H. and M.S. thank the Deutsche Forschungsgemeinschaft for their support via the collaborative research center SFB~652. 
		
\bibliography{biblio}

\end{document}